\documentclass[aps,prd,twocolumn,preprintnumbers,
superscriptaddress,showpacs,floatfix]{revtex4}

\usepackage{bm}
\usepackage{epsfig}
\usepackage{graphics}

\def\be{\begin{equation}}
\def\ee{\end{equation}}
\def\bea{\begin{eqnarray}}
\def\eea{\end{eqnarray}}

\renewcommand\({\left(}
\renewcommand\){\right)}

\newcommand\eq[1]{Eq.~(\ref{#1})}

\newcommand\mpl{M_{\rm P}}

\def\calp{{\cal P}}

\newcommand\sub[1]{_{\rm #1}}

\newcommand{\lsim}{\mbox{\raisebox{-.9ex}{~$\stackrel{\mbox{$<$}}{\sim}$~}}}
\newcommand{\gsim}{\mbox{\raisebox{-.9ex}{~$\stackrel{\mbox{$>$}}{\sim}$~}}}

\begin{document}

\preprint{UT-STPD-3/03}
\preprint{LANCS-16-07/03}

\title{Curvaton Dynamics}

\author{K. Dimopoulos}
\email{k.dimopoulos1@lancaster.ac.uk}
\affiliation{Physics Department, Lancaster University, 
Lancaster LA1 4YB, U.K.}
\affiliation{Department of Physics, University of Oxford, 
Keble Road, Oxford OX1 3RH, U.K.}
\author{G. Lazarides}
\email{lazaride@eng.auth.gr}
\affiliation{Physics Division, School of Technology,
Aristotle University of Thessaloniki,
Thessaloniki 54124, Greece}
\author{D. H. Lyth}
\email{d.lyth@lancaster.ac.uk}
\affiliation{Physics Department, 
Lancaster University, Lancaster LA1 4YB, U.K.}
\author{Roberto Ruiz de Austri}
\email{rruiz@gen.auth.gr}
\affiliation{Physics Division, School of Technology,
Aristotle University of Thessaloniki,
Thessaloniki 54124, Greece}

\date{\today}

\begin{abstract}
In contrast to the inflaton's case, the curvature perturbations due to the 
curvaton field depend strongly on the evolution of the curvaton before its 
decay. We study in detail the dynamics of the curvaton evolution during and 
after inflation. We consider that the flatness of the curvaton potential may 
be affected by supergravity corrections, which introduce an effective mass
proportional to the Hubble parameter. We also consider that the curvaton 
potential may be dominated by a quartic or by a non-renormalizable term. 
We find analytic solutions for the curvaton's evolution for all these 
possibilities. In particular, we show that, in all the above cases, the 
curvaton's density ratio with respect to the background density of the 
Universe decreases. Therefore, it is necessary that the curvaton decays only 
after its potential becomes dominated by the quadratic term, which results in
(Hubble damped)
sinusoidal oscillations. In the case when a non-renormalizable term dominates 
the potential, we find a possible non-oscillatory attractor solution that 
threatens to erase the curvature perturbation spectrum. Finally, we study the 
effects of thermal corrections to the curvaton's potential and show that, if 
they ever dominate the effective mass, they lead to premature thermalization 
of the curvaton condensate. To avoid this danger, a stringent bound has to be 
imposed on the coupling of the curvaton to the thermal bath.
\end{abstract}

\maketitle

\section{Introduction}

Observation of the  Cosmic Microwave Background Radiation
({\sc cmbr}) anisotropy has now confirmed that structure in the 
Universe originates as 
density perturbation, already  present a few Hubble times
before cosmological  scales enter the
horizon, with an almost flat  spectrum
\cite{wmapspergel,wmappeiris,wmapng,wmap}. 
The only known explanation  for this `primordial' density perturbation
is that it  originates, during an era of almost
exponential inflation, from the vacuum fluctuation of a 
light scalar field \cite{note1}.
The primordial density  perturbation is predominantly 
adiabatic in character \cite{wmappeiris,juan}, 
and hence characterized by a single 
quantity which may be taken to be  the spatial
curvature perturbation $\zeta$. According to observation, the spectrum
of the curvature perturbation (roughly its mean-square)  on
cosmological scales is \cite{wmapspergel}
\be
\calp_\zeta = \(  2\times 10^{-5} \)^2
\label{cobe}
\,.
\ee

 According to the usual hypothesis, the primordial
curvature perturbation
is generated solely by the inflaton field,  defined as
the one whose value determines the end of inflation. No other
field is involved, and the curvature perturbation is conserved
after inflation ends. As a result, the  spectrum of the primordial
curvature perturbation in this `inflaton scenario' is essentially
determined by the inflation model alone \cite{book,laza}. 
The only influence of the 
subsequent cosmology is to determine precisely the number $N\lsim 60$ of
$e$-folds of inflation after our Universe leaves the horizon.

It has been pointed out in Ref.~\cite{lw} (see also Refs.~\cite{sylvia,lm})
that the inflaton scenario is not the only possible way of generating large 
scale structure from inflation. The primordial density perturbation may 
instead originate from the vacuum fluctuation of some `curvaton' field 
$\sigma$, different from the inflaton field. The curvaton scenario has 
received a lot of attention 
\cite{dllr1,dl,luw,mt1,andrew,mt2,fy,hmy,hofmann,bck1,%
bck2,mormur,ekm,mwu,postma,fl,gl,kostas,giov,lu,mcdonald,ejkm,dlnr,%
ekm2,pm,kkt,lw03,lw032,lm1} because it opens up new possibilities both for
model-building and for  observation \cite{note2}.

In the  curvaton scenario, the  curvaton energy density is supposed to be 
initially negligible so that spacetime is practically unperturbed. It 
becomes significant only during some era of radiation domination, when the 
curvaton field is undergoing a (Hubble damped)
sinusoidal oscillation causing its energy 
density to grow relative to the radiation background. (In this paper, 
we take $\sigma$ to be a real scalar field.) While this happens,
the curvaton energy density $\rho_\sigma$ is proportional to the square of 
the amplitude $\bar\sigma$ of the oscillation, and the fractional perturbation
in both of these quantities is time-independent. Finally, the curvaton decays
before nucleosynthesis, leaving behind a conserved curvature perturbation 
which is proportional to the fractional energy density perturbation that 
existed before the curvaton decayed,
\begin{equation}
\zeta=\frac13 r\frac{\delta\rho_\sigma}{\rho_\sigma}=
\frac23 r\frac{\delta\bar\sigma}{\bar\sigma}
\label{spec}
\,.
\end{equation}
In the (quite good \cite{mwu,lm1}) approximation of sudden curvaton decay,
the constant $r$ is the energy density fraction of the curvaton at the decay
epoch,
\be
r\simeq\left.\frac{\rho_\sigma}{\rho}\right|\sub{dec}
\label{rdef}
\,.
\ee

The fractional perturbation in $\bar\sigma$ 
will be related to the fractional perturbation of the curvaton field
during inflation by some factor $q$,
\be
 \frac{\delta\bar\sigma}{\bar\sigma} 
  = q  \frac{\delta\sigma_*}{\sigma_*} \label{qdef}
\,,
\ee
where the subscript $*$ denotes the epoch when the cosmological scales exit 
the horizon during inflation. Using the well-known expression 
\mbox{$\calp_{\delta\sigma_*}^\frac12=H_*/2\pi$}, valid for any light scalar 
field, with $H$ the Hubble
parameter,  one obtains the prediction of the  curvaton scenario
\cite{lw,luw}
\be
\calp_\zeta^\frac12 = 
\frac2{6\pi} 
r q \frac{H_*}{\sigma_*} 
\label{spec1}
\,.
\ee
(We are ignoring, in this paper,  the possible scale-dependence of the 
curvature perturbation.)

This prediction  depends on the numbers
 $q$ and $r$, which encode the evolution of the curvaton
field 
between horizon exit during inflation and the epoch of curvaton
decay.
 The
 evolution is  given in terms of the effective potential $V(\sigma,H)$ by
the well-known equation
\be
\ddot\sigma + 3H\dot\sigma + V' =  0
\label{KG}
\,,
\ee
and its perturbation 
\be
\ddot {(\delta\sigma)} + 3H \dot {(\delta \sigma)} + V'' \delta \sigma =0
\label{dfKG}
\,,
\ee
where the prime and the dot denote differentiation with respect to $\sigma$
and the cosmic time $t$ respectively. 
The explicit dependence of the potential on $H$ comes from the 
time-dependent fields, which  must be present in the early Universe
in order to generate the energy density.

The simplest case is the one
where  the effective curvaton potential is so flat that the curvaton field 
actually has negligible evolution until the onset of (Hubble damped)
sinusoidal
oscillation. Then \mbox{$q=1$}, and $r$ may be calculated in terms of the
curvaton decay rate as described in Ref.~\cite{dlnr}.
This case may be expected to occur
\cite{lw,dlnr} if the curvaton is a pseudo-Nambu-Goldstone boson
({\sc pngb}),
so that it has a potential of the form
\be
V\simeq  (vm)^2 \( 1 - \cos\(\frac \sigma v \) \)
\label{pngb}
\,,
\ee
whose flatness is protected by a  global symmetry. 

In this paper, we consider the case  that the curvaton is not a {\sc pngb}.
Instead of \eq{pngb}, we adopt a form for the potential which allows
a range of possibilities,
\be
V(\sigma,H) = \frac12 \( m^2\!\pm cH^2\! + g^2 T^2 \)\sigma^2\!+ 
\frac{\lambda_n}{(n+4){\mbox{!}}}\frac{\sigma^{n+4}}{M^n}\,.
\ee
For the curvaton model to work $m$ must be far smaller than $H_*$.
The term \mbox{$\pm cH^2$} represents the effect of those fields to which
the curvaton has 
gravitational-strength couplings,
 on the assumption that supergravity is valid.
It may be positive or negative, and its value will change at the 
end of  inflation and again at reheating 
\cite{randall,randall2,thermal2}. Generically
\mbox{$c\sim 1$}, but during inflation one must have \mbox{$c_*\ll 1$} so that
the curvaton qualifies as a `light' field. The term $g^2T^2$ represents the 
effective coupling $g$ of the curvaton to those fields which are in
equilibrium at temperature $T$ after inflation.  
Finally, we have kept the leading 
term \mbox{$n\geq 0$} in the presumably infinite 
sum of quartic and higher terms. In the case of non-renormalizable terms
(\mbox{$n>0$}), we take $M$ to be 
 the ultra-violet cutoff of the effective field theory, 
say the Planck scale, and assume that  \mbox{$\lambda_n\lsim 1$}.

Using this form for the potential, we study the
 evolution of the curvaton field in a very general way. 
We verify that the curvaton contribution to
the energy density, if initially negligible,  remains so until the
(Hubble damped)
sinusoidal oscillation sets in, and we  also show how to calculate 
the numbers $r$ and $q$ which enter into the prediction for the curvature
perturbation. 

The plan of the paper is as follows. In Section~\ref{setup}, we lay out the
stage for our study. In particular, we provide the basic equations for the
background evolution of the Universe and we briefly introduce the curvaton 
model with emphasis on how the curvaton imposes its perturbations onto the
Universe and on what the requirements for a successful curvaton are.
Also, in this section, we discuss and justify the choice of the form of the 
curvaton's effective potential. In Section~\ref{durinf}, we study the 
evolution of the curvaton during inflation and estimate the value of the field
at the end of inflation as well as the amplitude of its perturbations at that 
time. In Section~\ref{afterinf}, we investigate the evolution of the curvaton
field and its density ratio with respect to the background density after the 
end of inflation. We find analytic solutions in all cases considered, which
describe both $\sigma(t)$ and $\rho_\sigma(t)$. In Section~\ref{curvdom},
we employ the results of the previous section to calculate, in all cases 
considered, the time when the curvaton comes to dominate the Universe or, if 
it decays earlier than this, the corresponding value of the density ratio $r$ 
at the time of decay. We also calculate in detail the value of $q$, relating 
the curvaton's fractional perturbation at decay with its initial fractional
perturbation, obtained in inflation. In Section~\ref{Tcorr}, we study the 
effects of temperature corrections to the effective potential on the evolution 
of the curvaton and find upper bounds on the curvaton's coupling $g$ 
to the thermal bath, which prevent these effects from being destructive.
In Section~\ref{exam}, we present a concrete model realization of our 
findings. Assuming a realistic set of values for the inflationary parameters,
we study the model in detail and, by enforcing all the constraints and 
requirements, we identify the parameter space for a successful curvaton. 
Finally, in Section~\ref{concl}, we discuss our results and 
present our conclusions.

In this paper, we use units such that \mbox{$c=\hbar=1$} so that Newton's 
gravitational constant is \mbox{$G^{-1}=8\pi m_P^2$}, where 
\mbox{$m_P=2.4\times 10^{18}$GeV} is the reduced Planck mass.

\section{The set up}\label{setup}

\subsection{The background evolution of the Universe}

The dynamics of the Universe expansion is determined by the Friedmann equation,
which, for a spatially flat Universe, reads
\begin{equation}
\rho=3(m_PH)^2,
\label{fried}
\end{equation}
where $\rho$ is the density of the Universe and \mbox{$H\equiv\dot{a}/a$} 
with $a$ being the scale factor of the Universe.

The dynamics of a real, homogeneous scalar field $\sigma$ in the expanding 
Universe is determined by the Klein-Gordon equation of motion given in 
Eq.~(\ref{KG}). It is well--known that a homogeneous scalar field can be 
treated as a perfect fluid with density and pressure given, respectively, by 
\begin{eqnarray}
\rho_\sigma\equiv\rho_{\rm kin}+V & \quad {\rm and} \quad &
p_\sigma\equiv\rho_{\rm kin}-V
\label{rp}
\end{eqnarray}
with \mbox{$\rho_{\rm kin}\equiv\frac{1}{2}\dot{\sigma}^2$} being its kinetic 
energy density. In view of this fact, we will model the content of the 
Universe as a collection of perfect fluids with equations of state of the 
form: \mbox{$\rho_i=w_ip_i$}, where $\rho_i$ and $p_i$ is the density and 
pressure of each fluid respectively and $w_i$ is the corresponding barotropic 
parameter. In particular, for matter \{radiation\}, we have \mbox{$w_m=0$} 
\{\mbox{$w_\gamma=\frac{1}{3}$}\}. In most of the history of
the Universe, the total density $\rho$ is dominated by one of the above 
components. 

The energy conservation (continuity) equation for the Universe, reads
\begin{equation}
\dot{\rho}=-3H(\rho+p)\,.
\label{energy}
\end{equation}
The above can be applicable to each of the component fluids individually,
if only they are independent so that their densities $\rho_i$ are 
conserved, diluted only by the expansion of the Universe according to 
Eq.~(\ref{energy}). 

If $w$ is the barotropic parameter of the dominant fluid then, 
for a constant $w$, Eq.~(\ref{energy}) gives \cite{note3}
\begin{equation}
\rho\propto a^{-3(1+w)}.
\label{raw}
\end{equation}
Using this and Eq.~(\ref{fried}), it is easy to find that, for 
\mbox{$w\neq -1$}, the Hubble parameter 
and the density of the Universe evolve in time as
\begin{eqnarray}
\hspace{-.3cm}
H(t)=\frac{2t^{-1}}{3(1+w)} & \;{\rm and}\; & \rho=\frac{4}{3(1+w)^2}
\left(\frac{m_P}{t}\right)^2,
\label{Hrhot}
\end{eqnarray}
from which it is evident that \mbox{$a\propto t^{2/3(1+w)}$}.
%

During inflation the Universe is dominated by the potential density of
another scalar field, the inflaton. According to this, Eq.~(\ref{rp})
suggests that the equation of state of the Universe during inflation
is \mbox{$w\simeq -1$}, which, in view of Eq.~(\ref{raw}),
suggests that \mbox{$\rho\simeq$ const.}. Hence (c.f. Eq.~(\ref{fried}))
the Hubble parameter remains roughly constant.

\subsection{The curvaton model}

The curvaton is a scalar field $\sigma$ other than the inflaton that may be 
responsible for the curvature perturbation in the Universe. One of the merits 
of considering such a field is the liberation of inflation model-building from 
the stringent requirements imposed by the observations of the Cosmic 
Background Explorer ({\sc cobe}). Indeed, the curvaton model changes the 
{\sc cobe} constraint on the energy scale of inflation $V_{\rm inf}^{1/4}$
into an upper bound \cite{dl}. Furthermore, in the context of the curvaton 
model, one can achieve a remarkably flat superhorizon spectrum of curvature 
perturbations, which is in agreement with the recent data from the 
Wilkinson Microwave Anisotropy Probe ({\sc wmap})
\cite{wmapspergel,wmappeiris}. Finally, since the curvaton (in contrast to the 
inflaton) is not related to the Universe dynamics during inflation it can be 
associated with much lower energy scales than the ones corresponding to 
inflation and, therefore, may be easily linked with lower energy (e.g. TeV) 
physics. This is why physics beyond the standard model provides many 
candidates for the curvaton, such as sfermions, string axions, 
or even the radion of large extra dimensions.

According to the curvaton scenario, the curvature perturbations, generated 
during inflation due to the quantum fluctuations of the inflaton field, are 
{\em not} the ones which cause the observed {\sc cmbr} anisotropy and seed the
formation of large scale structure. This is 
because their contribution is rendered negligible compared to the curvature 
perturbations introduced by another scalar field, the curvaton. In a similar 
manner as with the inflaton, the curvaton field receives an almost scale 
invariant, superhorizon spectrum of perturbations during inflation. However, 
at that time, the density of the curvaton field is subdominant, and the field 
lies frozen at some value displaced from the minimum of its potential. After 
the end of inflation, the field eventually unfreezes and begins oscillating. 
While doing so, the density fraction of the curvaton to the overall energy 
density increases. As a result, before decaying (or being thermalized) the 
curvaton may come to dominate (or nearly dominate) the Universe, thereby 
imposing its own curvature perturbation spectrum. After this, it decays into 
standard model particles, creating the thermal bath of the standard hot big 
bang.

\subsubsection{The curvature perturbation}

On a foliage of spacetime corresponding to spatially flat hypersurfaces, 
the curvature perturbation attributed to each of the Universe components is
given by \cite{luw}
\begin{equation}
\zeta_i\equiv-H\frac{\delta\rho_i}{\dot{\rho}_i}\,.
\label{zetai}
\end{equation}
Then the total curvature perturbation $\zeta(t)$, which also satisfies 
Eq.~(\ref{zetai}), may be calculated as follows.
Using that \mbox{$\delta\rho=\sum_i\delta\rho_i$} and Eq.~(\ref{energy}),
it is easy to find
\begin{equation}
(\rho+p)\zeta =\sum_i(\rho_i+p_i)\zeta_i\;.
\label{zeta}
\end{equation}
Now, since in the curvaton scenario all contributions to the curvature 
perturbation other than the curvaton's are negligible, we find that
\begin{equation}
\zeta_0
=\zeta_\sigma\left(\frac{1+w_\sigma}{1+w}\right)_{\rm dec}
\left.\frac{\rho_\sigma}{\rho}\right|_{\,\rm dec},
\label{z}
\end{equation}
where \mbox{$\zeta_0\equiv{\cal P}_\zeta^{\frac12}=2\times 10^{-5}$} is the 
curvature perturbation observed by {\sc cobe}, and $\zeta_\sigma$ is the 
curvature perturbation of the curvaton when the latter decays (or thermalizes).
From the above, it is evident that (c.f. Eq.~(\ref{rdef}))
\begin{equation}
r\equiv\frac{\zeta_0}{\zeta_\sigma}\simeq
\left.\frac{\rho_\sigma}{\rho}\right|_{\,\rm dec}.
\label{r}
\end{equation}

The curvaton decays when \mbox{$H\sim\Gamma$}, where
\begin{equation}
\Gamma=\max\{\Gamma_\sigma,\Gamma_T\}\,
\label{G}
\end{equation}
with $\Gamma_\sigma$ being the decay rate of the curvaton and $\Gamma_T$ 
the thermalization rate, corresponding to the thermal evaporation of the 
curvaton condensate. Note that, after the curvaton decay, the total curvature 
perturbation remains constant, i.e. 
\mbox{$\zeta_0=\zeta_{\rm dec}$}.

Suppose that the curvaton, just before decaying (or being thermalized),
is oscillating in a potential of the form \mbox{$V\propto\sigma^\alpha$}, where
\mbox{$\alpha$} is an even, positive number. Then, according to 
Ref.~\cite{turner}, the density of the oscillating field scales as 
\begin{equation}
\rho_\sigma\propto a^{-\frac{6\alpha}{\alpha+2}}\,, 
\label{ralpha}
\end{equation}
which means that \mbox{$\dot{\rho}=-\frac{6\alpha}{\alpha+2}\,\rho H$}. Using 
this, Eq.~(\ref{zetai}) gives
\begin{equation}
\zeta_\sigma=\frac{\alpha+2}{6}\;\left.\frac{\delta\bar\sigma}{\bar\sigma}
\right|_{\,\rm dec},
\label{zf}
\end{equation}
where we used that 
\mbox{$\delta\rho_\sigma/\rho_\sigma=\alpha\,\delta\sigma/\sigma$}.

Note also that, using Eq.~(\ref{energy}) for the oscillating curvaton, it is
easy to show that
\begin{equation}
w_\sigma=\frac{\alpha-2}{\alpha+2}={\rm const.}\quad.
\label{wf}
\end{equation}
Thus, in the case of a quadratic potential, the oscillating scalar field
behaves like pressureless matter, whereas for a quartic potential it behaves 
like radiation. For a (Hubble damped)
sinusoidal oscillation (\mbox{$\alpha=2$}), the above
give \mbox{$\zeta_\sigma=\frac23(\delta\bar\sigma/\bar\sigma)_{\rm dec}$} 
(c.f. Eq.~(\ref{spec})).

\subsubsection{The curvaton requirements}

Apart from providing the correct curvature perturbation, a successful curvaton 
needs to satisfy a number of additional requirements.
These can be outlined as follows:

\begin{itemize}
\item{\bf Masslessness:}
In order for the curvaton field to obtain a superhorizon spectrum of 
perturbations during inflation, at least on cosmological scales, the field 
needs to be effectively massless, when these scales exit the 
horizon. Thus, at that time, we require that
\begin{equation}
V''\ll H^2_*\;.
\label{mslsns}
\end{equation}

\item{\bf Misalignment:}
For the curvature perturbations due to the curvaton to be 
Gaussian, it is necessary that, when the cosmological scales exit the horizon
during inflation, the curvaton field is significantly displaced from the 
minimum of its potential. Thus, we require that
\begin{equation}
(\sigma-\sigma_{\rm min})_*
> H_*\;.
\label{mslnmnt}
\end{equation}

\item{\bf WMAP:}
According to the recent {\sc wmap} observations, to avoid excessive 
non-Gaussianity in the density perturbation spectrum, the curvaton 
needs to decay (or thermalize) when the density ratio $r$ of Eq.~(\ref{r})
is sufficiently large and satisfies the bound \cite{wmapng}
\begin{equation}
r>9\times 10^{-3}.
\label{wmap}
\end{equation}
Obviously, in order for the field not to be thermalized too soon 
after the end of inflation, we require that its coupling $g$ to the thermal 
bath, generated by the inflaton's decay, is small enough. 

\item{\bf Nucleosynthesis:}
Due to the above {\sc wmap} bound on $r$, one has to demand that
the curvaton should have decayed (and not just be thermalized) by the time 
when big bang nucleosynthesis ({\sc bbn}) takes place (at temperature 
\mbox{$T_{\rm BBN}\sim 1$ MeV}), which imposes the constraint
\begin{equation}
\Gamma_\sigma>H_{\rm BBN}\sim 10^{-24}{\rm GeV}\,.
\label{bbn}
\end{equation}

\end{itemize}

\subsection{The effective potential}

The form of the perturbative scalar potential is 
\cite{note4}
\begin{equation}
V(\sigma)=\frac{1}{2}m^2\sigma^2+\sum_{n\geq 0}
\frac{\lambda_n}{(n+4){\mbox{!}}}\frac{\sigma^{n+4}}{M^n}\,,
\label{Vpert}
\end{equation}
where \mbox{$0<\lambda_n\lsim 1$} and $M$ is some large mass-scale 
(e.g. $m_P$) determining when the non-renormalizable terms become 
important. We will consider that the sum is dominated by the term of the
($n+4$)$-$th order. We will also take into account temperature corrections due
to the possible coupling $g$ of the field with an existing thermal bath of 
temperature $T$. Finally, we will consider corrections to the 
scalar potential due to supergravity 
effects. These corrections introduce an effective mass, whose magnitude is 
determined by the Hubble parameter $H(t)$ \cite{randall2}. All in all, the 
form of our scalar potential is
\begin{equation}
V(\sigma)=\frac{1}{2}(m^2\pm c H^2+g^2T^2)\sigma^2+
\frac{\lambda_n}{(n+4){\mbox{!}}}\frac{\sigma^{n+4}}{M^n}\,,
\label{V0}
\end{equation}
where \mbox{$0<c,g\lsim 1$} \cite{note5}. The masslessness requirement demands 
that, at least during inflation, \mbox{$c_*\ll 1$}. This may be due to some 
(approximate) symmetry or due to accidental cancellations of specific 
K\"{a}hler corrections, which means that it depends on the particular curvaton 
model \cite{note6}. Obviously, the 
masslessness requirement also demands \mbox{$m\ll H_*$} during inflation.

Depending on the sign in front of $c$, the above potential
may have a global minimum or a local maximum at the origin. In the latter
case and when \mbox{$\sqrt{c}H\gg m,gT$}, the potential is characterized by 
global minima at $\pm\sigma_{\rm min}$ such that
\begin{equation}
\sigma_{\rm min}(H)=
\left[(n+4)\mbox{!}\,
\frac{cH^2M^n}{\lambda_n}\right]^{\frac{1}{n+2}}
\hspace{-.2cm}\sim (cH^2M^n)^{\frac{1}{n+2}}\,,
\label{fmin}
\end{equation}
where, in the last equation, we have absorbed $\lambda_n/(n+4)$! into $M$ 
(for \mbox{$n\rightarrow 0$}, we can set \mbox{$M^n\rightarrow 4$!$/\lambda$}).
In this case, in order to ensure that the potential remains positive
we may add a constant \mbox{$V_0\gsim cH^2\sigma_{\rm min}^2$} \cite{note7}.

If the local minimum is displaced from the origin then one can express
the potential in terms of 
\be
\hat{\sigma}\equiv\sigma-\sigma_{\rm min}\,,
\label{fhat}
\ee
in which case, for \mbox{$\sqrt{c}H\gg m,gT$}, the potential of Eq.~(\ref{V0}) 
is recast as
\begin{equation}
V(\hat{\sigma})=\frac{1}{2}(n+2)cH^2\hat{\sigma}^2+
\frac{\lambda_n}{M^n}
\sum_{k=0}^{n+1}\alpha_k\sigma_{\rm min}^{n+1-k}\hat{\sigma}^{3+k},
\label{linear}
\end{equation}
where \mbox{$\alpha_k^{-1}\equiv$ ($n+1-k$)! ($3+k$)!}. Since the series is 
finite, the above is an exact expression and not an approximation.
It is evident that all the terms of the sum in Eq.~(\ref{linear}) are 
comparable to each-other when \mbox{$\hat{\sigma}\simeq\sigma_{\rm min}$} and 
the sum reduces to $\sigma_{\rm min}^{n+4}\sum_{k=0}^{n+1}\alpha_k$. For 
\mbox{$\hat{\sigma}\gg\sigma_{\rm min}$}, however, the sum is dominated by the
term of the largest order which is \mbox{$\hat{\sigma}^{n+4}/(n+4)$!}. 
Obviously, when \mbox{$\hat{\sigma}\ll\sigma_{\rm min}$}, only the 
quasi-quadratic 
(`quasi' in the sense that \mbox{$H=H(t)$}) term is important. Thus, in all 
cases, the potential can be roughly approximated by
\begin{equation}
V(\sigma)\sim cH^2\hat{\sigma}^2+\frac{\sigma^{n+4}}{M^n}\,,
\label{V}
\end{equation}
where we considered that, when \mbox{$\hat{\sigma}\gg\sigma_{\rm min}$}, then
\mbox{$\hat{\sigma}\simeq\sigma$}.

In the case of $+c$ in 
Eq.~(\ref{V0}), \mbox{$\sigma_{\rm min}=0$} and \mbox{$\hat{\sigma}=\sigma$}.
In the $-c$ case, \mbox{$\sigma_{\rm min}\neq 0$} but the complicated 
structure near the local maximum is felt only when the field is near 
\mbox{$\hat{\sigma}\sim\sigma_{\rm min}$}. Since we attempt a generic study of 
the behaviour of the field's dynamics, Eq.~(\ref{V}) will suffice for 
\mbox{$\sqrt{c}H\gg m,gT$}. It is true, however, that
a different higher-order term may dominate for different values of the field. 
This is one of the reasons for treating $n$ as a free parameter here.
In general, we expect that the order of the dominant higher--order term 
increases with $\sigma$. This can be understood as follows.

Reinstating the $\lambda_n$ coefficients, it is easy to see that the ratio of 
two higher--order terms of different orders is
\begin{equation}
\frac{V_n}{V_{n+k}}\sim\frac{\lambda_n}{\lambda_{n+k}}
\left(\frac{M}{\sigma}\right)^k.
\label{ls}
\end{equation}
This means that, if all the $\lambda_n$'s are of the same magnitude then, for 
\mbox{$\sigma\ll M$}, the sum of the higher--order terms in 
Eq.(\ref{Vpert}) will be dominated by the lowest order (quartic term), while,
for \mbox{$\sigma\gg M$}, it will be the highest order that dominates. Since 
the sum is infinite, this means that $V(\sigma\gg M)$ blows up and the 
perturbative approximation is no longer valid. However, if the $\lambda_n$ 
coefficients are not of the same magnitude, then it is possible that the sum 
of the higher--order terms is dominated by a single term, higher than the 
quartic, before blowing up. Increasing $\sigma$ may change the order of this 
term to an even higher one for given $\lambda_n$'s as shown by Eq.~(\ref{ls}).

\section{During inflation}\label{durinf}

\subsection{Regimes of evolution}

During inflation, our scalar field is expected to roll down its potential 
toward the minimum. This roll may be of three distinct types, depending on 
the initial conditions of the field or, more precisely, on where exactly the 
field lies originally in its potential. Below, we briefly discuss these 
regimes assuming that, for all practical purposes, the Hubble parameter during 
inflation is almost constant. We also assume, for simplicity, that the minimum
of the potential lies at the origin (if not one should simply set 
\mbox{$\sigma\rightarrow\hat{\sigma}$}) and consider the potential of 
Eq.~(\ref{V}). Without loss of generality we take \mbox{$\sigma>0$}.

\begin{itemize}
\item{\bf The fast-roll regime:}
The field is in this regime when it lies at a region of its potential
such that \mbox{$V''\gg H_*^2$}. Due to the masslessness requirement
during inflation, we expect \mbox{$c_*\ll 1$}. Thus, the field may be in the
fast-roll regime only if it lies into the higher--order part of its 
potential. Then it is easy to see that the border of the fast-roll regime
corresponds to \mbox{$V''(\sigma_{\rm fr})\sim H_*^2$}, where
\begin{equation}
\sigma_{\rm fr}\sim(H_*^2M^n)^{\frac{1}{n+2}}\sim 
c_*^{-\frac{1}{n+2}}\sigma^*_{\rm min}
\label{fFR}
\end{equation}
with \mbox{$\sigma^*_{\rm min}\equiv\sigma_{\rm min}(H_*)$}.

Once in the fast-roll regime (i.e. for \mbox{$\sigma>\sigma_{\rm fr}$}),
the field rapidly rolls until it reaches $\sigma_{\rm fr}$. After this, the 
motion of the field is no longer underdamped but feels instead the drag of
the Universe expansion. Any kinetic energy assumed during the fast-roll
will be depleted very fast after crossing $\sigma_{\rm fr}$
as can be easily seen from Eq.~(\ref{KG}) \cite{note8}.

\item{\bf The slow-roll regime:}
This regime corresponds to \mbox{$V''\ll H_*^2$}. The field's motion is 
overdamped by the Universe expansion and Eq.~(\ref{KG})
can be approximated as
\begin{equation}
3H\dot{\sigma}\simeq-V'(\sigma)\,,
\label{SR}
\end{equation}
which is the familiar form of the inflaton's equation of motion in slow-roll 
inflation. 

\item{\bf The quantum regime:}
Classically, the field will continue to slow-roll until it reaches 
the minimum of its potential. However, an effectively massless scalar field 
during inflation undergoes particle production by generating superhorizon 
perturbations due to its quantum fluctuations. In that respect, {\em the 
quantum fluctuations may affect the classical motion of the field if the 
generation of the perturbations becomes comparable to the slow-roll motion of 
the field}. Let us elaborate a bit on this issue.

In a region of fixed size $H_*^{-1}$ (horizon), the effect of the quantum 
fluctuations can be represented \cite{stoch} as a random walk with step given
by the Hawking temperature \mbox{$\delta\sigma=T_H\equiv H_*/2\pi$} per Hubble 
time. The slow-roll motion is comparable to this ``quantum kick'' when 
\mbox{$\delta\sigma\sim\dot{\sigma}/H_*$}, or 
equivalently when \mbox{$\sigma\simeq\sigma_Q$}, where 
$\dot{\sigma}$ is given by Eq.~(\ref{SR})
and $\sigma_Q$ is implicitly defined by the condition
\begin{equation}
V'(\sigma_Q)\equiv H_*^3\;.
\label{QR}
\end{equation}
(Here $\sigma$ corresponds to region of fixed size $H_*^{-1}$). 
One can understand the above also as follows. The energy density 
corresponding to the quantum fluctuations, which exit the horizon, is 
\mbox{$\rho_{\rm qm}\sim T_H^4\sim H_*^4$}
\cite{note9}. The quantum regime begins when this
energy density is comparable to the kinetic density of the slow-roll
\mbox{$\rho_{\rm kin}=\frac{1}{2}\dot{\sigma}^2$}. Comparing the two, one 
finds the same condition, Eq.~(\ref{QR}), for the border of the quantum regime 
$\sigma_Q$ \cite{note10}. 

After the onset of the quantum regime, the coherent motion of $\sigma_*$ 
ceases. Instead, the field configuration spreads toward the origin by random 
walk. 
As a result, the mean value at horizon exit of the curvaton field in our 
Universe (denoted by $\sigma_*$) may lie anywhere within the range 
\mbox{$|\sigma_*|\leq\sigma_Q$}, which means that, typically, its value would 
be \mbox{$\sigma_*\sim\sigma_Q$}. The same is true if the field finds itself 
already into the quantum regime at the onset of inflation.

\end{itemize}

Thus, from the above, we can sketch the typical evolution of the field during 
inflation. Since the curvaton is expected to have negligible energy density 
when the cosmological scales exit the horizon during inflation, if we demand 
that initially \mbox{$\rho_\sigma\sim V_{\rm inf}$}, this means that the field 
must originally lie in the steep (and curved) part of its potential (where it 
cannot act as a curvaton) and, therefore, it will engage into fast-roll. Very 
soon, however, fast-roll sends the field below $\sigma_{\rm fr}$ and the 
curvaton enters its slow-roll regime. Whether it remains there until the end 
of inflation or not depends on the duration of the inflationary period. If 
inflation lasts long enough, the field will slow-roll down to $\sigma_Q$, 
where it will be ``stabilized'' by the action of its quantum fluctuations. The 
quantum drift may further reduce $\sigma_*$ somewhat but, since in the quantum 
regime the field is oblivious of the potential $V$, there is no real 
motivation to drag it towards the origin. Still, \mbox{$\sigma_*\ll\sigma_Q$} 
can be selected anthropically, i.e. by assuming that we happen to live at a 
special place in the Universe.

\subsection{Initial conditions at the end of inflation}

The above enable us to speculate on the initial conditions of the curvaton 
after the end of inflation. Indeed, given sufficiently long inflation, we
expect \mbox{$\sigma_{\rm end}\sim\sigma_Q$} and 
\mbox{$\dot{\sigma}_{\rm end}\sim 0$} because the coherent motion of the field
ceases once it enters into the quantum regime. However, if inflation ends 
before the field reaches the quantum regime, we have 
\mbox{$\sigma_{\rm end}>\sigma_Q$}, where the 
subscript `end' denotes the end of inflation. Thus, we see that
\begin{equation}
\sigma_Q\lsim\sigma_{\rm end}<\sigma_{\rm fr}\;.
\label{fend}
\end{equation}

Using Eq.~(\ref{QR}) and also that \mbox{$V'\sim\sigma V''$} for a 
perturbative potential, it is easy to see that
\begin{equation}
\sigma_Q\sim\left(\frac{H_*^2}{V''}\right)H_*\gg H_*\;.
\end{equation}
Since, in all cases, \mbox{$\sigma\geq\sigma_{\rm end}\gsim\sigma_Q$}, we see 
that {\em the misalignment requirement for the curvaton is naturally 
guaranteed by masslessness}. 

Let us estimate $\sigma_Q$ using Eq.~(\ref{V}). Whether $\sigma_Q$ lies in the 
quasi-quadratic part of the potential or not depends on how large is the 
value of $c_*$. Indeed it is easy to see that
\begin{equation}
\sigma_Q\sim\left\{
\begin{array}{lr}
H_*/c_* & c_0<c_*<1\\
H_*/c_0\sim(H_*^3M^n)^{\frac{1}{n+3}} & c_*\leq c_0
\end{array}\right.,
\label{fQ}
\end{equation}
where
\begin{equation}
c_0\equiv(H_*/M)^{\frac{n}{n+3}}.
\label{c0}
\end{equation}
From the above, since \mbox{$c_0,c_*\ll 1$} during inflation, it is evident 
that the misalignment requirement is, quite generically, satisfied.
Using Eqs.~(\ref{fFR}), (\ref{fQ}) and (\ref{c0}), it is easy to show that,
when \mbox{$c_*\leq c_0$},
\begin{equation}
\frac{\sigma_Q}{\sigma_{\rm fr}}\sim c_0^{\frac{1}{n(n+2)}}<1
\end{equation}
and, therefore, fast-roll always ends before reaching the quantum regime.

We should point out here that $c_*$ during inflation cannot be arbitrarily 
small. Indeed, due to the coupling $g$ between the curvaton and other fields, 
one expects a contribution of the form \mbox{$g^2T_H^2$} to the effective mass 
during inflation, which introduces the bound \mbox{$c_*\geq g^2$}. Thus, 
during inflation,
\begin{equation}
g^2\leq c_* \ll 1\,.
\label{crange}
\end{equation}
Note, also, that, if \mbox{$c_*<(m/H_*)^2$}, 
the soft mass $m$ dominates the quasi-quadratic term of the potential. 

In the $-c$ case, the above are valid with the substitution 
\mbox{$\sigma\rightarrow\hat{\sigma}$}. If \mbox{$c_*\leq c_0$} then 
\mbox{$\sigma_Q\simeq\hat{\sigma}_Q\gg\sigma_{\rm min}$}. However, if
\mbox{$c_0<c_*\ll 1$} then \mbox{$\hat{\sigma}_Q\ll\sigma_{\rm min}$}, which 
means that \mbox{$\sigma_Q\sim\sigma_{\rm min}$} in this case.

\subsection{The amplitude of perturbations}

We can now estimate the amplitude of the curvaton's almost scale invariant 
superhorizon spectrum of perturbations. The evolution of $\delta\sigma$ after 
horizon crossing is determined by Eq.~(\ref{dfKG}),
which, for \mbox{$V''\ll H_*^2$} and 
\mbox{$0\lsim\dot{(\delta\sigma)}_0<H_*^2$}, 
gives
\begin{equation}
\delta\sigma\simeq\delta\sigma_0\,
\exp\left[-\frac{1}{3}\left(\frac{V''}{H_*^2}\right)H_*\Delta t\right]\sim
H_*e^{-\tilde{\eta}\Delta N},
\label{frz0}
\end{equation}
where \mbox{$\delta\sigma_0=H_*/2\pi$} at horizon exit,
\mbox{$\tilde{\eta}\equiv V''/3H_*^2$} and \mbox{$\Delta N=H_*\Delta t$} is 
the elapsing e-foldings of inflation. For the cosmological scales, 
\mbox{$\Delta N\lsim 60$} by the end of inflation. Thus, for 
\mbox{$V''/H_*^2<0.05$}, we see that 
\mbox{$\delta\sigma\approx\delta\sigma_0\sim H_*$}, i.e. the perturbation 
remains frozen. Thus, in view also of Eq.~(\ref{fQ}), for the amplitude of the 
perturbation spectrum we find
\begin{equation}
\left.\frac{\delta\sigma}{\sigma}\right|_{\rm end}\sim
\frac{H_*}{\sigma_{\rm end}}
\;\lsim\;\frac{H_*}{\sigma_Q}\sim\max\{c_*,c_0\}\,.
\label{df/fend}
\end{equation}

Here we should mention the fact that, while \mbox{$\tilde{\eta}\neq 0$} 
and \mbox{$\dot{H}_*\neq 0$} during inflation leave the amplitude of the 
perturbation spectrum largely unaffected, they do affect the slope of the 
spectrum, causing departure from scale invariance. Indeed, in Ref.~\cite{luw} 
it is shown that the spectral index of the curvature perturbation spectrum is 
given by
\begin{equation}
n_s=1+2(\tilde{\eta}-\epsilon)\,,
\label{ns}
\end{equation}
where \mbox{$\epsilon\equiv-(\dot{H}/H^2)_*\ll 1$}.

\section{After the end of inflation}\label{afterinf}

After the end of inflation, the inflaton begins oscillating
around its vacuum expectation value. These coherent oscillations
correspond to massive particles (inflatons), whose energy density dominates 
the Universe. Thus, after inflation ends the Universe enters a matter 
dominated period. The inflatons eventually decay creating a thermal 
bath of temperature $T$. 
The Universe becomes radiation dominated only after 
the 
decay of the inflaton field (reheating), which takes place at 
\mbox{$H_{\rm reh}\sim\Gamma_{\rm inf}$}, where $\Gamma_{\rm inf}$ is the 
decay rate of the inflaton field and the subscript `reh' corresponds to 
reheating. The temperature at this moment is the so-called reheat temperature 
$T_{\rm reh}$. After reheating, the Universe enters a radiation dominated 
period. 

In the following, we will use Eq.~(\ref{KG}) to follow the evolution of the 
curvaton field after the end of inflation. At the first stage, we will assume 
that \mbox{$\sqrt{c}H\gg m,gT$} so that the potential may be approximated by
Eq.~(\ref{V}). Because, after the end of inflation,
there are no perturbations generated due to quantum fluctuations to inhibit 
the field's motion, the curvaton evolves entirely classically, following 
Eq.~(\ref{KG}). Its evolution depends on which of the two terms in 
Eq.~(\ref{V}) is the dominant. Below we study both cases individually.

\subsection{The quasi-quadratic case}\label{qq}

Suppose that it is the quasi-quadratic term that dominates the potential 
in Eq.~(\ref{V}). Then we can write
\begin{equation}
V(\sigma)\simeq\frac12\; cH^2(t)\sigma^2,
\label{Vquad}
\end{equation}
where, for simplicity, we have dropped the hat on $\sigma$. Inserting the
above into Eq.~(\ref{KG}) and substituting \mbox{$dt=(1+w)td\tau$}, we obtain
\begin{equation}
\frac{\partial^2\sigma}{d\tau^2}+(1-w)\frac{d\sigma}{d\tau}+\frac{4c}{9}\sigma
=0\,.
\label{ftau}
\end{equation}

The above admits oscillating solutions if $c$ is larger that $c_{\rm x}$, where
\begin{equation}
\sqrt{c_{\rm x}}\equiv\frac{3}{4}(1-w)\,.
\label{cx}
\end{equation}

\subsubsection{Case $c> c_{\rm x}$}

Assuming negligible initial kinetic density and
after some algebra, it can be shown that, in this case, the solution for 
$\sigma(t)$ is
\begin{eqnarray}
 & \sigma(t) & = \sigma_0
\left(\frac{t}{t_0}\right)^{-\frac{1}{2}\left(\frac{1-w}{1+w}\right)}
\left\{
\cos\left[\frac{2\sqrt{c-c_{\rm x}}}{3(1+w)}
\ln\left(\frac{t}{t_0}\right)\right]
\right.+\nonumber\\
 & & \nonumber\\
& & +\left.\frac{3(1-w)}{4\sqrt{c-c_{\rm x}}}
\sin\left[\frac{2\sqrt{c-c_{\rm x}}}{3(1+w)}\ln\left(\frac{t}{t_0}\right)\right]
\right\}.
\label{solu1}
\end{eqnarray}
It is evident from the above that, when \mbox{$c\rightarrow c_{\rm x}$}, the 
oscillation frequency reduces to zero.

Since \mbox{$V\propto H^2\sigma^2\propto (\sigma/t)^2$} and 
\mbox{$\rho_{\rm kin}\propto\dot{\sigma}^2\propto(\sigma/t)^2$}
and in view of Eq.~(\ref{Hrhot}), we find that 
\begin{equation}
\rho_\sigma
\propto a^{-\frac{3}{2}(3+w)}.
\label{rfquad1}
\end{equation}
Comparing the above with Eq.~(\ref{raw}), we obtain
\begin{equation}
\frac{\rho_\sigma}{\rho}\propto a^{-\frac{3}{2}(1-w)},
\label{ratio1}
\end{equation}
which means that the density ratio of the curvaton is decreasing
and cannot lead to curvaton domination.

Note here that, although during inflation masslessness requires 
\mbox{$c_*\ll 1$}, this is not necessarily so after inflation, when $c$ may 
assume a different value. In fact, unless the (approximate) symmetry that 
enables masslessness persists after the end of inflation, we would expect 
\mbox{$c\sim 1$} and, therefore, \mbox{$c>c_{\rm x}$} is quite plausible.

\subsubsection{Case $c\leq c_{\rm x}$}

\begin{figure}[t]
\includegraphics[width=75mm,angle=0]{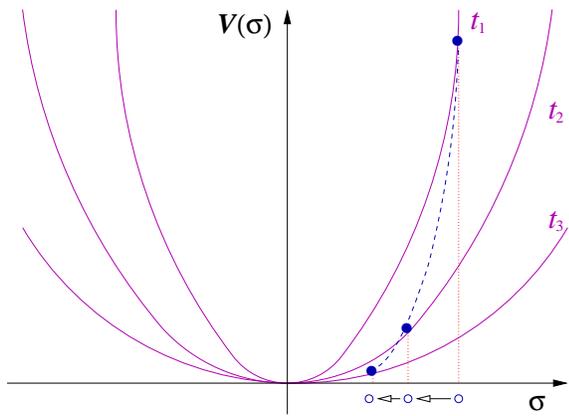}
\caption{\label{scaling} 
Illustration of the scaling solution in the quasi-quadratic case, where
\mbox{$m_{\rm eff}\sim\sqrt{c}H(t)$}. For \mbox{$t_1<t_2<t_3$}, the curvaton
field $\sigma$ rolls down a potential which changes in time so that the 
field's value gradually reduces towards $\sigma_{\rm min}$.}
\end{figure}

In this case, Eq.~(\ref{ftau}) does not admit oscillating solutions. Instead,
for \mbox{$c<c_{\rm x}$} and negligible initial kinetic density,
one finds the scaling 
solution (see Fig.~\ref{scaling})
\begin{eqnarray}
 & \sigma(t) & = \sigma_0
\left(\frac{t}{t_0}\right)^{-\frac{1}{2}\left(\frac{1-w}{1+w}\right)}\times
\nonumber\\
 & & \times\left[
\left(1+\frac{3(1-w)}{4\sqrt{c_{\rm x}-c}}\right)
\left(\frac{t}{t_0}\right)^{\frac{2\sqrt{c_{\rm x}-c}}{3(1+w)}}
+\right.\nonumber\\
 & & \nonumber\\
& & +\left.
\left(1-\frac{3(1-w)}{4\sqrt{c_{\rm x}-c}}\right)
\left(\frac{t}{t_0}\right)^{-\frac{2\sqrt{c_{\rm x}-c}}{3(1+w)}}\,\right].
\label{solu2}
\end{eqnarray}

The second term in the square brackets soon becomes negligible.
Again \mbox{$V\propto\rho_{\rm kin}\propto(\sigma/t)^2$}, which means 
that~\cite{note11}
\begin{equation}
\rho_\sigma
\propto a^{-\frac{3}{2}(3+w)+2\sqrt{c_{\rm x}-c}}.
\label{rfquad2}
\end{equation}

From the above, it is evident that, when \mbox{$c\rightarrow c_{\rm x}$}
the scaling of $\rho$ reduces to the one given by Eq.~(\ref{rfquad1}).
Using Eq.~(\ref{raw}), we now find for the curvaton's density ratio
\begin{equation}
\frac{\rho_\sigma}{\rho}\propto a^{-\frac{3}{2}(1-w)+2\sqrt{c_{\rm x}-c}}.
\label{ratio2}
\end{equation}
The above again shows that the curvaton's density ratio is decreasing and 
cannot lead to curvaton domination. To make this clearer, consider that, for 
\mbox{$c\rightarrow 0$}, Eq.~(\ref{cx}) suggests that 
\mbox{$\rho_\sigma/\rho\rightarrow$ const.}, which means that, for all 
\mbox{$c>0$}, the exponent in Eq.~(\ref{ratio2}) is negative.

The solution for \mbox{$c=c_{\rm x}$}, with negligible initial kinetic density,
is easily found to be
\begin{equation}
\sigma(t)=\sigma_0\left[1+\frac{1}{2}\left(\frac{1-w}{1+w}\right)
\ln\left(\frac{t}{t_0}\right)\right]
\left(\frac{t}{t_0}\right)^{-\frac{1}{2}\left(\frac{1-w}{1+w}\right)}.
\end{equation}
Ignoring the logarithmic time dependence, the above results in the same scaling
for $\rho_\sigma$ as in Eq.~(\ref{rfquad1}).

From Eqs.~(\ref{solu1}) and (\ref{solu2}), we see that 
\mbox{$\delta\sigma/\sigma=$ const.} whether the field is oscillating or not.
In that sense was Eq.~(\ref{solu2}) called a `scaling' solution. Thus, 
in view also of Eq.~(\ref{zf}), in the quasi-quadratic case we find
\begin{equation}
\left.\frac{\delta\sigma}{\sigma}\right|_{\rm end}\simeq
\left.\frac{\delta\sigma}{\sigma}\right|_{\rm dec}\sim\;\zeta_\sigma\;.
\label{dff1}
\end{equation}

\subsection{The higher--order case}\label{nonren1}

This case corresponds to 
\mbox{$\hat{\sigma}\simeq\sigma\gg\sigma_{\rm min}$}, when
the scalar potential in Eq.~(\ref{V}) can be approximated as
\begin{equation}
V(\sigma)\simeq\frac{\sigma^{n+4}}{M^n}\,.
\label{Vnonren}
\end{equation}

In contrast to the previous case, where the effective mass was determined by 
the Hubble parameter, in this case the roll of the field does not necessarily
commence immediately after the end of inflation. Indeed, since at least during 
the latest stage of inflation (when the cosmological scales exited the horizon 
and afterwards) the masslessness requirement demands \mbox{$V''\ll H$},
just after the end of inflation, the field originally lies in the 
potential at a point where its motion is largely overdamped \cite{note12}. 
Consequently, the field remains frozen at the value $\sigma_{\rm end}$. One 
can easily show this as follows. Considering a negligible initial kinetic 
density, the motion of the field is determined by Eq.~(\ref{SR}), which gives
\begin{equation}
\left(\frac{\sigma_0}{\sigma}\right)^{n+2}\hspace{-.2cm}
\simeq 1+\frac{1}{9(1+w)}
\left(\frac{n+2}{n+3}\right)\left(1-\frac{H^2}{H_0^2}\right)
\frac{V''_0}{H^2}\,.
\end{equation}
Thus, as long as \mbox{$V''_0\ll H^2$}, we see that 
\mbox{$\sigma\approx\sigma_0$}, i.e. the field remains frozen. 
However, while $H(t)$ decreases there will be a moment 
when \mbox{$V''(\sigma_{\rm end})\sim H$} after which the field unfreezes and 
begins to roll again down its potential. Note that, as expected, a similar 
result applies to the perturbation of the field $\delta\sigma$. For example,
solving Eq.~(\ref{dfKG}) for the matter era following the end of inflation,
one finds
\begin{eqnarray}
\delta\sigma & = & \delta\sigma_0\left(\frac{H}{H_0}\right)\left\{
\cos\left[\frac{2}{3}\frac{\sqrt{V''_0}}{H}\left(1-\frac{H}{H_0}\right)\right]+
\right.\nonumber\\
 & & \left.+\frac{3}{2}\frac{H_0}{\sqrt{V''_0}}
\sin\left[\frac{2}{3}\frac{\sqrt{V''_0}}{H}\left(1-\frac{H}{H_0}\right)\right]
\right\},
\label{frz}
\end{eqnarray}
which, in the regime \mbox{$\sqrt{V''_0}\ll H(t)\ll H_0$}, becomes
\mbox{$\delta\sigma\approx\delta\sigma_0[1+(\frac{H}{H_0})]
\approx\delta\sigma_0$}.
Thus, for the amplitude of the perturbation spectrum we expect
\begin{equation}
\left.\frac{\delta\sigma}{\sigma}\right|_{\rm osc}\simeq
\left.\frac{\delta\sigma}{\sigma}\right|_{\rm end},
\label{df/fosc}
\end{equation}
where the subscript `osc' denotes the onset of the $\sigma$ oscillations.

As demonstrated in Ref.~\cite{turner}, the energy density of an oscillating
scalar field in an expanding Universe with potential 
\mbox{$V\propto\sigma^{n+4}$} scales as
\begin{equation}
\rho_\sigma\propto a^{-6\left(\frac{n+4}{n+6}\right)}.
\label{rnonren}
\end{equation}
However, when the scalar field unfreezes, it does not always engage into 
oscillations. Indeed, it is easy to show that Eq.~(\ref{KG}) with a scalar 
potential of the form of Eq.~(\ref{Vnonren}) has the following exact 
solution:
\begin{equation}
\sigma_{\rm solu}(t)\!=\!
\left[2M^n\frac{(1-w)(n+4)-4}{(n+4)(n+2)^2(1+w)}\right]^{\frac{1}{n+2}}
\!\!t^{-\frac{2}{n+2}}.
\label{solu}
\end{equation}

Of course, in order for the system to follow this solution, it has to have the 
correct initial conditions. However, in Ref.~\cite{attr}, a stability analysis 
has shown that the above solution is an attractor to the system for any
\mbox{$n>n_c$}, where 
\begin{equation}
n_c\equiv 2\left(\frac{1+3w}{1-w}\right).
\label{nc}
\end{equation}

Attractor solutions are lethal for the curvaton scenario. Indeed, one of the
principle characteristics of attractor solutions is that they are insensitive
to initial conditions. In other words, if the system is to follow an attractor
solution, all the memory of the initial conditions will be eventually lost.
This is a disaster for the curvaton because it means that the superhorizon 
spectrum of perturbations will be erased. To understand this better, consider 
the above solution $\sigma_{\rm solu}(t)$. If this is an attractor solution 
then, after the field unfreezes, it will tend to approach this solution, i.e. 
\mbox{$\sigma(t)\rightarrow\sigma_{\rm solu}(t)$}. However, if we consider a 
perturbed value of the field $\tilde{\sigma}=\sigma+\delta\sigma$ then again 
\mbox{$\tilde{\sigma}(t)\rightarrow\sigma_{\rm solu}(t)$}, which means that
\mbox{$\delta\sigma\rightarrow 0$}. Thus, the perturbation spectrum 
$(\delta\sigma/\sigma)_{\rm end}$, present at the end of inflation, will 
rapidly diminish so that \mbox{$(\delta\sigma/\sigma)_{\rm dec}\rightarrow 0$} 
and, therefore, from Eq.~(\ref{zf}), we see that 
\mbox{$\zeta_\sigma\rightarrow 0$}.

Consequently, in order to avoid the above catastrophe, we have to enforce
the {\em attractor constraint} :
\begin{equation}
n<n_c\;,
\label{attr}
\end{equation}
i.e. we forbid the curvaton field to lie at the end of inflation deep enough 
in the higher--order part of its potential so that the attractor constraint 
is violated. 

The attractor constraint is more stringent in 
the matter dominated era (before reheating) because then \mbox{$n_c=2$}. Thus, 
if the curvaton unfreezes before reheating is completed then it should not lie
into the higher--order part of $V$ beyond quartic terms. Similarly, 
for the radiation era, \mbox{$n_c=6$} so the constraint is relaxed up to 
terms of order 8, if the field is to unfreeze after reheating. 

Of course, one can imagine that, even if the initial value after the end of 
inflation does lie in the forbidden regime, after unfreezing there is a chance 
that, while approaching the attractor, the field manages to roll down to
parts of its potential where the attractor constraint is not violated before
its perturbation spectrum becomes entirely erased. Although this is, in 
principle, marginally possible it does not result in anything more
than effectively making the borders of the forbidden zone somewhat ``fuzzy''.
Thus, in the following, we will disregard this possibility and consider the 
attractor constraint as strict. 

The attractor constraint has a number of consequences on the evolution of the 
field. For example, from Eqs.~(\ref{Vnonren}) and (\ref{rnonren}), one finds
for the amplitude $\bar{\sigma}$ of the oscillations
\begin{eqnarray}
\bar{\sigma}\propto a^{-\frac{6}{n+6}} & \Rightarrow & 
\frac{\bar{\sigma}}{\sigma_{\rm min}}\propto
H^{\frac{2}{n+6}\left(\frac{n_c+6}{n_c+2}-\frac{n+6}{n+2}\right)},
\label{barf}
\end{eqnarray}
where we used also Eqs.~(\ref{fmin}) and (\ref{nc}). The above shows that the 
attractor constraint in Eq.~(\ref{attr}) guarantees that the ratio 
$\bar{\sigma}/\sigma_{\rm min}$ is increasing and, therefore, the 
oscillation amplitude will never drop as low as $\sigma_{\rm min}$. This means 
that once the higher--order term is dominant it will continue to be so 
without ever the quasi-quadratic term in Eq.~(\ref{V}) taking over. One can 
also show this as follows. The effective mass for the potential of 
Eq.~(\ref{Vnonren}) is \mbox{$m_{\rm eff}^2\equiv V''\sim\sigma^{n+2}/M^n$}. 
Then, using the above, one can show that
\begin{eqnarray}
m_{\rm eff}\propto a^{-3\left(\frac{n+2}{n+6}\right)} & \Rightarrow & 
m_{\rm eff}\propto
H^{\left(\frac{n+2}{n+6}\right)/\left(\frac{n_c+2}{n_c+6}\right)},
\label{meff}
\end{eqnarray}
from which it is evident that, due to the attractor constraint,
 $m_{\rm eff}^2$ decreases slower than $cH^2$ and, therefore, the 
quasi-quadratic term in Eq.~(\ref{V}) will never take over once the field 
finds itself in the higher--order part of its potential at the end of 
inflation.

From Eqs.~(\ref{raw}) and (\ref{rnonren}), we find that the density ratio
of the oscillating curvaton scales as
\begin{equation}
\frac{\rho_\sigma}{\rho}\propto a^{-6\left(\frac{n+4}{n+6}\right)+3(1+w)}.
\label{ratio3}
\end{equation}
It is easy to see that the exponent in the above is non-positive for
\mbox{$n\geq-2(\frac{1-3w}{1-w})$}, which is valid both in the matter and 
the radiation eras for all \mbox{$n\geq 0$}. Thus we conclude that the
oscillating curvaton in the higher--order case is unable to dominate the 
Universe, as in the quasi-quadratic case. In fact, we have found that the 
scaling of the curvaton's density is such that, for \mbox{$0\leq n<n_c$}, it
is decreasing slower than the quasi-quadratic case but not as slow as the 
background density of the Universe.

It is easy to show that the solution in Eq.~(\ref{solu}) scales as
\begin{equation}
\rho_\sigma(\sigma_{\rm solu})\propto a^{-3(1+w)\left(\frac{n+4}{n+2}\right)}.
\end{equation}
Comparing this with Eq.~(\ref{rnonren}), one can see that the scaling coincides
in the border case, when \mbox{$n=n_c$}. As shown in Ref.~\cite{attr}, in 
this border case, even though the energy density of the scalar field does scale
according to the attractor solution, the field engages into oscillations.
However, as we show in Sec.~\ref{qsec}, its perturbation spectrum is not
preserved and so \mbox{$n=n_c$} cannot be allowed.

For the oscillating curvaton, we have \mbox{$\delta\sigma/\sigma=$ const.},
since both $\sigma$ and $\sigma+\delta\sigma$ undergo the same scaling given by
Eq.~(\ref{barf}). This fact, in view also of Eqs.~(\ref{zf}) and 
(\ref{df/fosc}), suggests that
\begin{equation}
\left.\frac{\delta\sigma}{\sigma}\right|_{\rm end}\sim\;\zeta_\sigma\;,
\label{dff2}
\end{equation}
which is the same result as in the quasi-quadratic case 
(c.f. Eq.~(\ref{dff1})). We elaborate more on this point in Sec.~\ref{qsec}.

Therefore, in both the quasi-quadratic and in the higher--order cases, we 
have shown that {\em there is no significant damping of the 
curvaton's curvature perturbation after the end of inflation}
\cite{note13}.
Consequently, using Eq.~(\ref{df/fend}), we find the range for the amplitude 
of the curvature perturbations of $\sigma$:
\begin{equation}
\zeta_0\leq\zeta_\sigma\lsim\max\{c_*,c_0\}\,.
\label{zrange}
\end{equation}

We have also shown that, in both the quasi-quadratic and in the 
higher--order 
cases, the curvaton is unable to dominate or even to increase its density 
fraction with respect to the background density of the Universe. This means
that for the curvaton scenario to work {\em the soft mass m should take over 
before the curvaton decays} (or thermalizes). In Sec.~\ref{curvdom}, we 
investigate this further.

\subsection{\boldmath The $-c$ case revisited}\label{qq-c}

Before concluding this section some reference to the $-c$ case in 
Eq.~(\ref{V0}) is called upon. Due to Eq.~(\ref{barf}), a negative sign in 
front of $c$ has no effect on the dynamics of the higher--order case, 
because the quasi-quadratic term remains always subdominant. In contrast, it
may have an important effect in the quasi-quadratic case.

The minus in front of $c$ may exist in inflation and turn into a plus after
inflation ends. This has no other effect than setting the value of 
$\sigma_{\rm end}$ to 
\begin{equation}
\sigma_{\rm end}=\hat{\sigma}_{\rm end}+\sigma_{\rm min}^{\rm end}
\approx\sigma_{\rm min}^{\rm end}\;,
\label{fend-c}
\end{equation}
where \mbox{$\sigma_{\rm min}^{\rm end}\equiv\sigma_{\rm min}(H_{\rm end})$}.
The above is so because, in the quasi-quadratic case,
\mbox{$\hat{\sigma}_{\rm end}<\sigma_{\rm min}$}.
Another possibility is that the minus existed there both before and after the 
end of inflation, or that it appeared after the end of inflation. Again
Eq.~(\ref{fend-c}) applies \cite{note14}. 

However, now that the quasi-quadratic term, after 
the end of inflation, has a negative sign, the evolution of the field may be 
affected. Let us study this issue.

In this case, Eq.~(\ref{Vquad}) becomes
\begin{equation}
V(\sigma)\simeq\frac12\; cH^2(t)(\sigma-\sigma_{\rm min})^2.
\label{Vquad2}
\end{equation}
Substituting this into Eq.~(\ref{KG}), we can recast it as
\begin{equation}
\ddot{\sigma}+3H(t)\dot{\sigma}+cH^2(t)\sigma=cH^2(t)\;\sigma_{\rm min}(t)\,,
\label{KG2}
\end{equation}
where we see that there is now a source term on the right-hand-side 
({\sc rhs}). The solution to the above equation, therefore, is of the form
\begin{equation}
\sigma(t)=\sigma_{\rm hom}(t)+\sigma_{\rm par}(t)\,,
\label{fhompec}
\end{equation}
where the homogeneous solution $\sigma_{\rm hom}(t)$ is given in 
Eq.~(\ref{solu1}) or Eq.~(\ref{solu2}) depending on the value of $c$. 
It is easy to see that the particular solution 
$\sigma_{\rm par}(t)$ is of the form
\begin{equation}
\sigma_{\rm par}(t)=A\;\sigma_{\rm min}(t)\,,
\label{fpec}
\end{equation}
where, from Eq.~(\ref{fmin}),
\mbox{$\sigma_{\rm min}(t)\propto t^{-\frac{2}{n+2}}$} and
$A$ is the constant
\begin{equation}
A=\left[1+\frac{9}{c}\left(\frac{1+w}{n+2}\right)
\left(\frac{1+w}{n+2}-1\right)\right]^{-1}.
\label{A}
\end{equation}
According to Eq.~(\ref{fend-c}), the initial conditions, in this case, suggest 
that \mbox{$\sigma_{\rm end}\simeq\sigma_{\rm min}^{\rm end}$}. Thus, 
$\sigma_0$ in Eq.~(\ref{solu1}) or Eq.~(\ref{solu2}) for the 
$\sigma_{\rm hom}(t)$ is 
\mbox{$\sigma_0\simeq(1-A)\sigma_{\rm min}^{\rm end}$}, because
initially \mbox{$\sigma_{\rm}(t_0)=A\,\sigma_{\rm min}^{\rm end}$}. 
Therefore, for \mbox{$t\gg t_{\rm end}$}, the solution is
\begin{eqnarray}
\sigma(t) & \sim & (1-A)\sigma_{\rm min}^{\rm end}
\left(\frac{t}{t_{\rm end}}\right)^{-\frac{1}{2}\left(\frac{1-w}{1+w}\right)
\left(1-\sqrt{1-\hat{c}/c_{\rm x}}\;\right)}+\nonumber\\
& & +A\;\sigma_{\rm min}^{\rm end}
\left(\frac{t}{t_{\rm end}}\right)^{-\frac{2}{n+2}},
\label{solu3}
\end{eqnarray}
where \mbox{$\hat{c}\equiv$ min\{$c,c_{\rm x}$\}} and, in the oscillating case
(i.e. when \mbox{$c>c_{\rm x}$}), $\sigma(t)$ corresponds to the amplitude of 
the oscillations. 

It is interesting to note that the attractor constraint in Eq.~(\ref{attr})
results in
\begin{eqnarray}
\hspace{-.5cm}n<n_c\!\Rightarrow & \hspace{-.2cm}-\frac{2}{n+2}\!<\! 
-\frac{1}{2}\left(\frac{1-w}{1+w}\right)
\left(1-\sqrt{1-\hat{c}/c_{\rm x}}\right)
 & .
\label{exponents}
\end{eqnarray}
Therefore, we see that, {\em if the attractor constraint is satisfied, the 
particular solution decreases faster than the homogeneous and, hence, it 
soon becomes negligible.}
If this is the case then there is no modification to the dynamics described in 
Sec.~\ref{qq} other than the initial condition, which is the one given in
Eq.~(\ref{fend-c}).

However, since the field is not in the higher--order part of its 
potential, there is no compelling reason for the attractor constraint to be 
satisfied. Indeed, for sufficiently large $n$, the particular solution 
decreases less fast than the homogeneous. When this is so, the domination 
of the particular solution results in 
\mbox{$\sigma(t)\propto\sigma_{\rm min}$}. In 
other words, for sufficiently large $n$, $\sigma_{\rm par}$ takes over and 
keeps the field at a constant ratio \mbox{$A>1$} to $\sigma_{\rm min}$. 

The problem is that, in contrast to the homogeneous solution, {\em the 
particular solution does not carry the memory of the initial conditions 
and, therefore, does not preserve the perturbation spectrum of the field}.
This means that the domination of the particular solution should be avoided,
which imposes the constraint
\begin{equation}
n+2<(n_c+2)
\left(1-\sqrt{1-\min\left\{1,\frac{c}{c_{\rm x}}\right\}}\;\right)^{-1}.
\label{nmin}
\end{equation}
The above constraint is more relaxed than the attractor constraint of 
Eq.~(\ref{attr}) and becomes even more so the smaller $c$ is.

One final point to note before we conclude this section is that, from 
Eqs.~(\ref{dff1}) and (\ref{fend-c}), in the quasi-quadratic case with 
negative sign, after the end of inflation the curvature perturbation of the 
field is
\begin{equation}
\zeta_\sigma\sim
\left.\frac{H}{\sigma_{\rm min}}\right|_{\rm end}\ll c_*\ll 1\,.
\label{zetaminus}
\end{equation}
We elaborate more on this point in Sec.~\ref{smallq}.

\section{Curvaton domination}\label{curvdom}

We now turn our attention to the evolution of the curvaton's density ratio
over the background density \mbox{$\rho_\sigma/\rho$}, which will determine 
when the curvaton will dominate the Universe, or, if it decays 
(or thermalizes) earlier, what is the value of $r$ of Eq.~(\ref{r}). 

The history of the Universe may be quite complicated. There may be more than
one periods of inflation, some of them lasting only a limited number of 
e-foldings (e.g. thermal inflation). There may also be more than one fields 
acting as curvatons, in the sense that they obtain a superhorizon spectrum of 
perturbations during inflation, which they manage, at some point, to impose 
onto the Universe. This is not improbable since the field content of
theories beyond the standard model is rich in potential curvatons. If this 
is the case, then the curvature perturbation of the Universe, that we observe 
today, corresponds 
e.g. to the curvaton which dominated the Universe last. (However, note that, if
the curvaton that decays last only nearly dominates the Universe, there may be
other significant contributions to the total curvature perturbation e.g. from
other, earlier curvatons.) Being aware of these possible complications, 
we do not pretend to describe the evolution of the actual history of the 
Universe. Instead, we provide the tools to study this history. To demonstrate 
the use of those tools, we implement them in a minimalistic scenario, where 
there is one, long period of inflation and one curvaton field.
In this minimalistic scenario, the inflationary period is followed by a 
matter era, dominated by the oscillating inflaton, and a radiation era,
commencing after reheating. We still ignore temperature corrections by 
assuming a negligible $g$ postponing their study to Sec.~\ref{Tcorr}.


As mentioned in the 
previous section, the curvaton has a chance to dominate only if it decays 
(or thermalizes) after the soft mass $m$ dominates the effective mass 
\mbox{$m_{\rm eff}\equiv\sqrt{V''}$}. Thus, we require
\begin{equation}
H_m>\Gamma\,,
\end{equation}
where $H_m$ is the Hubble parameter when \mbox{$m\sim m_{\rm eff}$},
which is different in the quasi-quadratic case and the higher--order case.
Below we treat both cases individually again, as in the previous section.

\subsection{The quasi-quadratic case}\label{qq1}

From Eqs.~(\ref{rfquad1}) and (\ref{rfquad2}) and in view also of 
Eq.~(\ref{Hrhot}), we find
\begin{equation}
\frac{\rho_\sigma}{\rho}\propto 
H^K,
\end{equation}
where
\begin{equation}
K\equiv\left(\frac{1-w}{1+w}\right)
\left(1-\sqrt{1-\min\left\{1,\frac{c}{c_{\rm x}}\right\}}\;\right).
\label{K}
\end{equation}
From the above and Eq.~(\ref{cx}), it is evident that
the exponent $K$ depends on whether we are in the matter or the radiation era. 
Indeed, for the matter and radiation eras respectively, we have
\begin{eqnarray}
K_{\rm MD} & \equiv & 1-\sqrt{1-\min\{1,16c/9\}}\label{KMD}\\
\hspace{-1.7cm}{\rm and}\hspace{1.7cm} & & \nonumber\\
K_{\rm RD} & \equiv & 
\frac{1}{2}\left(1-\sqrt{1-\min\{1,4c\}}\,\right),
\label{KRD}
\end{eqnarray}
which results in
\begin{eqnarray}
\left.\frac{\rho_\sigma}{\rho}\right|_{\rm MD} & \sim & 
\left(\frac{H}{H_{\rm end}}\right)^{K_{\rm MD}}
\;\left.\frac{\rho_\sigma}{\rho}\right|_{\rm end}
\label{ratioMD}\\
\hspace{-2.3cm}{\rm and}\hspace{2.3cm} & & \nonumber\\
\left.\frac{\rho_\sigma}{\rho}\right|_{\rm RD} & \sim & 
\left(\frac{H}{\Gamma_{\rm inf}}
\right)^{K_{\rm RD}}
\;\left.\frac{\rho_\sigma}{\rho}\right|_{\rm reh}.
\label{ratioRD}
\end{eqnarray}
Note that, for \mbox{$c\ll 1$}, we have
\begin{eqnarray}
K_{\rm MD}\;\stackrel{^{c\ll 1}}{\longrightarrow}\;\frac{8}{9}\,c
 & \quad{\rm and}\quad &
K_{\rm RD}\;\stackrel{^{c\ll 1}}{\longrightarrow}\; c\,.
\label{Kc0}
\end{eqnarray}
Thus, for \mbox{$c\rightarrow 0$}, the ratio $\rho_\sigma/\rho$ remains 
constant as expected according to the discussion after Eq.~(\ref{ratio2}). 

Using Eqs.~(\ref{fried}), (\ref{fQ}) and (\ref{Vquad}), we find
\begin{equation}
\left.\frac{\rho_\sigma}{\rho}\right|_{\rm end}\sim
c\left(\frac{\sigma_{\rm end}}{m_P}\right)^2\gsim
\frac{1}{c}\left(\frac{H_{\rm end}}{m_P}\right)^2
\label{ratioend}
\end{equation}
and
\begin{equation}
\left.\frac{\rho_\sigma}{\rho}\right|_{\rm reh}\sim
\left(\frac{\Gamma_{\rm inf}}{H_{\rm end}}\right)^{K_{\rm MD}}
\left.\frac{\rho_\sigma}{\rho}\right|_{\rm end}.
\label{ratioreh}
\end{equation}
Now, in the quasi-quadratic case, it is easy to see that
\begin{equation}
H_m\sim m/\sqrt{c}\,,
\label{Hmqq}
\end{equation}
Using this and Eqs.~(\ref{ratioMD}) and (\ref{ratioRD}), we obtain
\begin{eqnarray}
\left.\frac{\rho_\sigma}{\rho}\right|_{H\sim H_m} & \sim & 
c\left(\frac{\sigma_{\rm end}}{m_P}\right)^2
\left(\frac{m/\sqrt{c}}{H_{\rm end}}\right)^{K_{\rm MD}}\times\nonumber\\
 & & \times\;\min\left\{1,\frac{m/\sqrt{c}}{\Gamma_{\rm inf}}
\right\}^{K_{\rm RD}-K_{\rm MD}}.
\label{ratiomquad1}
\end{eqnarray}

When $H(t)$ falls below $H_m$, the quasi-quadratic term becomes dominated by
the quadratic term corresponding to the soft mass $m$ in Eq.~(\ref{V0}).
If \mbox{$c<1$} then there is a regime where something curious happens.
Indeed, for \mbox{$m<H(t)<m/\sqrt{c}$}, the scalar potential is dominated
by the quadratic term, for which \mbox{$V''=m^2\ll H^2$}, i.e. the 
field becomes overdamped and, consequently, freezes \cite{note15}. This does 
not happen in the case where the field was oscillating before $H_m$ was 
reached, because in this case \mbox{$c>c_{\rm x}\sim 1$}. However, it happens 
when the field is following the scaling solution in Eq.~(\ref{solu2}), 
according to which, until $H_m$ was reached, the field was rolling steadily 
towards the origin. After $H_m$, this roll is halted. The important point to 
be stressed here is that, roughly when $H_m$ is reached, 
{\em the field stops regardless of its value}. That is the whole configuration
of $\sigma$ freezes at the same moment and, therefore, 
{\em the perturbation spectrum is unaffected by the freezing}. The reason is 
evident: the possible freezing has to do with the comparison of $m$ and 
$m_{\rm eff}(t)$ 
with each-other and with $H(t)$. None of these quantities, in 
the quasi-quadratic case, is related to the actual value of $\sigma$ and, 
therefore, the freezing of the field configuration is $\sigma$-independent.

If the field configuration does freeze then, in the interval
\mbox{$m<H<H_m$}, the scaling of the curvaton's density ratio is only 
due to the decrease of $\rho$. Thus, from Eq.~(\ref{raw}), in this interval, 
we have
\begin{equation}
\frac{\rho_\sigma}{\rho}\propto a^{3(1+w)},
\label{ratiofrz}
\end{equation}
which is, for the first time, {\em increasing}. Is it possible for the 
curvaton to dominate (or nearly dominate) during this interval? 
Well, since \mbox{$\rho_\sigma/\rho\propto\rho^{-1}\propto H^{-2}$}, we find
\begin{equation}
\left.\frac{\rho_\sigma}{\rho}\right|_{H\sim m}\sim
\frac{1}{c}\left.\frac{\rho_\sigma}{\rho}\right|_{H\sim H_m}.
\label{ratiomquad2}
\end{equation}
Thus, after the whole interval of freezing has passed the amplification of the
density ratio is only a factor of $1/c$. Hence, at the end of the freezing 
interval, the curvaton's density ratio is
\begin{equation}\hspace{-1.8cm}
\left.\frac{\rho_\sigma}{\rho}\right|_m\!\!\!\!\sim\!
\left(\frac{\sigma_{\rm end}}{m_P}\right)^{\!\!2}\!\!
\left(\frac{m/\sqrt{c}}{H_{\rm end}}\right)^{\!\!K_{\rm MD}}
\hspace{-.3cm}\min\!\left\{1,\frac{m/\sqrt{c}}{\Gamma_{\rm inf}}
\right\}^{\!K_{\rm RD}\!-\!K_{\rm MD}}\hspace{-.15cm},\hspace{-1cm}
\label{ratiomquad}
\end{equation}
where we have set 
\mbox{$(\rho_\sigma/\rho)_m\equiv(\rho_\sigma/\rho)_{H\sim m}$}.
Notice now that, in the {\sc rhs} of the above,
all the factors are smaller  than (or at most equal to) unity, while all the 
exponents are positive. This means that the density ratio of 
Eq.~(\ref{ratiomquad}), even though somewhat increased compared to 
Eq.~(\ref{ratiomquad1}), is still expected to 
be much smaller than unity. It is hardly possible, therefore, that this ratio 
is able to satisfy the constraint of Eq.~(\ref{wmap}), let alone lead to 
domination. Therefore, curvaton domination will have to wait for the quadratic 
oscillations to begin. The initial density ratio for the curvaton at the onset 
of these final quadratic oscillations is given by Eq.~(\ref{ratiomquad}).

An important comment to be made here, regarding Eq.~(\ref{ratiomquad}), is that
the curvaton's density ratio (with respect to the background density at a 
given time) strongly depends on the value of $c$. Indeed, when 
\mbox{$c\ll 1$}, 
Eq.~(\ref{Kc0}) suggests that the two last factors on the {\sc rhs} of 
Eq.~(\ref{ratiomquad}) are negligible. On the other hand, if \mbox{$c\sim 1$}
then these factors are not negligible but, instead, they may strongly suppress
the density ratio. Indeed, it is evident that the curvaton's density ratio,
in this case, is suppressed at least by a factor of $(m/H_{\rm end})$, which
can be very small. The reason is that, for \mbox{$c\ll 1$}, the effective mass 
is small and the field is overdamped. On the other hand, if $c$ is comparable 
to unity both the quasi-quadratic oscillations and the scaling solution
substantially diminish the density of the curvaton. In fact, it is possible 
that this suppression is so efficient that the final quadratic regime is 
unable to counteract it. Hence, we have shown that {\em the quasi-quadratic 
evolution disfavours curvatons which admit substantial supergravity 
corrections to their effective potential}. This strengthens the case of the 
curvaton being a {\sc pngb} as discussed in Ref.~\cite{dlnr}. The above are 
more clearly demonstrated in the example analysed in Sec.~\ref{exam}.

%

\subsection{The higher--order case}

As discussed in Sec.~\ref{nonren1}, in this case, the field unfreezes and 
begins oscillating when \mbox{$m_{\rm eff}\sim H$}. Suppose, at first, 
that, between the onset of the field's oscillation and the time when the soft
mass $m$ dominates $m_{\rm eff}$, the dominant fluid component of the Universe
remains the same so that $w$ is unchanged. Then, using Eq.~(\ref{meff}), it is
easy to find that
\begin{equation}
H_m\sim\left(\frac{m}{H_{\rm osc}}
\right)^{\left(\frac{n+6}{n+2}\right)
\left(\frac{n_c+2}{n_c+6}-\frac{n+2}{n+6}\right)}m\,,
\label{Hmnonren}
\end{equation}
where we used that \mbox{$m_{\rm eff}(H_{\rm osc})\sim H_{\rm osc}$}.
Due to the attractor constraint in Eq.~(\ref{attr}) we see that the exponent
in the above is always positive. This means that 
\begin{equation}
H_{\rm osc}\geq m\quad\Rightarrow\quad H_m\leq m\,.
\label{Hmm}
\end{equation}
The above shows that, {\em the oscillation of the field 
will continue uninterrupted when the quadratic term takes over.}
This is because, for a pure quadratic potential, the field oscillations
begin when \mbox{$H(t)\sim \sqrt{V''}=m$}. Also, since 
\mbox{$H_{\rm osc}\geq H_m$} (for the higher--order oscillations to take 
place), the condition \mbox{$H_m\leq m$} is guaranteed if
\mbox{$H_{\rm osc}\leq m$}. Therefore, in all cases \mbox{$H_m\leq m$} and
{\em there is no intermediate freezing of the field once it unfreezes after 
the end of inflation}.

Now, employing Eqs.~(\ref{nc}), (\ref{ratio3}) and (\ref{Hmnonren}), we obtain
\begin{equation}
\left.\frac{\rho_\sigma}{\rho}\right|_m\sim
\left(\frac{m}{H_{\rm osc}}\right)^{\frac{(1-w)(n+6)-4}{n+2}}
\left(\frac{\sigma_{\rm end}}{m_P}\right)^2,
\label{ratiomnonren}
\end{equation}
where we also used
\begin{equation}
\left.\frac{\rho_\sigma}{\rho}\right|_{\rm osc}\sim
\left(\frac{\sigma_{\rm end}}{m_P}\right)^2,
\label{ratioosc}
\end{equation}
which is evident in view of Eq.~(\ref{fried}) and considering that
\mbox{$\rho_\sigma\sim m_{\rm eff}^2\sigma^2$} with 
\mbox{$m_{\rm eff}(H_{\rm osc})\sim H_{\rm osc}$} and
\mbox{$\rho_\sigma^{\rm osc}=\rho_\sigma^{\rm end}$}.

From the above, it is straightforward to find the curvaton's density ratio
at the onset of the quadratic oscillations in the case where this occurs
before reheating (i.e. \mbox{$H_m>\Gamma_{\rm inf}$}). Indeed, from 
Eq.~(\ref{ratiomnonren}), one gets
\begin{equation}
\left.\frac{\rho_\sigma}{\rho}\right|_m^{_{\rm MD}}\sim
\left(\frac{m}{H_{\rm osc}}\right)
\left(\frac{\sigma_{\rm end}}{m_P}\right)^2,
\label{ratiomnonrenMD}
\end{equation}
corresponding to (c.f. Eq.~(\ref{Hmnonren}))
\begin{equation}
H_m\sim\left(\frac{m}{H_{\rm osc}}
\right)^{\frac{1}{2}\left(\frac{2-n}{n+2}\right)}m\,.
\label{HmnonrenMD}
\end{equation}
Note that, for unfreezing before reheating, the attractor constraint demands
\mbox{$n<2$}. Similarly, we obtain the curvaton's density ratio at the 
onset of the quadratic oscillations in the case when unfreezing occurs
after reheating (i.e. \mbox{$H_{\rm osc}<\Gamma_{\rm inf}$}). This time,
Eq.~(\ref{ratiomnonren}) gives
\begin{equation}
\left.\frac{\rho_\sigma}{\rho}\right|_m^{_{\rm RD}}\sim
\left(\frac{m}{H_{\rm osc}}\right)^{\frac{2n}{3(n+2)}}
\left(\frac{\sigma_{\rm end}}{m_P}\right)^2,
\label{ratiomnonrenRD}
\end{equation}
corresponding to (c.f. Eq.~(\ref{Hmnonren}))
\begin{equation}
H_m\sim\left(\frac{m}{H_{\rm osc}}
\right)^{\frac{6-n}{3(n+2)}}m\,.
\label{HmnonrenRD}
\end{equation}
Note that, for unfreezing after reheating, the attractor constraint
is relaxed to \mbox{$n<6$}.

What happens, however, if \mbox{$H_m<\Gamma_{\rm inf}<H_{\rm osc}$}?
In this case, during the oscillations and before the quadratic term dominates,
reheating is completed and the Universe switches from being matter dominated 
to being radiation dominated. We may obtain $H_m$ and $(\rho_\sigma/\rho)_m$
working in the same manner as above. 

Employing Eq.~(\ref{meff}) both for the matter and radiation eras, we get
\begin{equation}
m\sim\left[\left(\frac{H_m}{\Gamma_{\rm inf}}\right)^{3/2}
\left(\frac{\Gamma_{\rm inf}}{H_{\rm osc}}\right)^2\right]^{\frac{n+2}{n+6}}
H_{\rm osc}\;.
\end{equation}
Solving for $H_m$ we obtain
\begin{equation}
H_m\sim
\left(\frac{m}{H_{\rm osc}}
\right)^{\frac{2}{3}\left(\frac{2-n}{n+2}\right)}
\left(\frac{m}{\Gamma_{\rm inf}}\right)^{1/3}m\,.
\label{HmnonrenMDRD}
\end{equation}
Since unfreezing occurs before reheating, the attractor constraint
demands \mbox{$n<2$}, which suggests that \mbox{$H_m\leq m$}
if \mbox{$H_{\rm osc}\geq m$} and \mbox{$m\leq\Gamma_{\rm inf}$}. Since
\mbox{$H_{\rm osc}\geq H_m$}, the condition \mbox{$H_m\leq m$} is 
automatically satisfied if \mbox{$H_{\rm osc}\leq m$}. Also, if 
\mbox{$m\geq\Gamma_{\rm inf}$} then again \mbox{$H_m\leq m$} is satisfied 
because \mbox{$H_m<\Gamma_{\rm inf}$} in
the first place. Therefore \mbox{$H_m\leq m$} in all cases, which means that
{\em the oscillations continue uninterrupted after the quadratic term 
dominates}. 

Now, using Eq.~(\ref{ratio3}), we get
\begin{eqnarray}
\left.\frac{\rho_\sigma}{\rho}\right|_{\rm reh} & \sim &
\left(\frac{\Gamma_{\rm inf}}{H_{\rm osc}}\right)^{2\left(
\frac{n+2}{n+6}\right)}
\left.\frac{\rho_\sigma}{\rho}\right|_{\rm osc},\\
\left.\frac{\rho_\sigma}{\rho}\right|_m\; & \sim &
\;\left(\frac{H_m}{\Gamma_{\rm inf}}\right)^{\frac{n}{n+6}}
\left.\frac{\rho_\sigma}{\rho}\right|_{\rm reh}.
\end{eqnarray}
Putting these together and also using Eqs.~(\ref{ratioosc}) and 
(\ref{HmnonrenMDRD}), we find
\begin{equation}
\left.\frac{\rho_\sigma}{\rho}\right|_m^{_{{\rm MD}\rightarrow{\rm RD}}}
\!\!\!\!\sim
\left(\frac{m}{H_{\rm osc}}\right)^{\frac{2n}{3(n+2)}}
\left(\frac{\Gamma_{\rm inf}}{H_{\rm osc}}\right)^{2/3}
\left(\frac{\sigma_{\rm end}}{m_P}\right)^2.
\label{ratiomnonrenMDRD}
\end{equation}

Summing up, in the above we have obtained the value of $H_m$ 
[Eqs.~(\ref{HmnonrenMD}), (\ref{HmnonrenRD}) and (\ref{HmnonrenMDRD})] and of 
the ratio $(\rho_\sigma/\rho)_m$ [Eqs.~(\ref{ratiomnonrenMD}), 
(\ref{ratiomnonrenRD}) and (\ref{ratiomnonrenMDRD})] in the cases
\mbox{$\Gamma_{\rm inf}\leq H_m\leq H_{\rm osc}$},
\mbox{$H_m\leq H_{\rm osc}\leq\Gamma_{\rm inf}$} and
\mbox{$H_m\leq\Gamma_{\rm inf}\leq H_{\rm osc}$} respectively.
We have found that the attractor constraint ensures that \mbox{$H_m\leq m$}
in all cases and, therefore, guarantees that {\em the oscillations of the 
field continue uninterrupted after the quadratic term due to the soft mass $m$ 
dominates the effective mass}.

Before moving on to the final quadratic oscillations, we provide an estimate
for $H_{\rm osc}$. Indeed, since 
\mbox{$H_{\rm osc}^2\sim m_{\rm eff}^2(\sigma_{\rm end})\sim
\sigma_{\rm end}^{n+2}/M^2$}, Eqs.~(\ref{fend}) and (\ref{fQ}) suggest that
\begin{equation}
H_{\rm osc}\gsim\sqrt{c_0}\,H_{\rm end}\;.
\label{Hosc}
\end{equation}

\subsection{The final quadratic oscillations}

When \mbox{$H(t)<H_m$}, the scalar potential becomes dominated by the 
quadratic term due to the soft mass
\begin{equation}
V(\sigma)\simeq\frac12\; m^2\sigma^2.
\label{Vm}
\end{equation}
According to Eq.~(\ref{wf}), a scalar field coherently oscillating in a 
quadratic potential corresponds to a collection of massive particles at rest
(curvatons in our case) and behaves like pressureless matter, i.e. 
\mbox{$\rho_\sigma\propto a^{-3}$}. Therefore, from Eq.~(\ref{raw}), one finds
\begin{equation}
\frac{\rho_\sigma}{\rho}\propto a^{3w}\propto H^{-\frac{2w}{1+w}}.
\label{ratio4}
\end{equation}
The above shows that, in the quadratic regime the curvaton's density ratio
is constant in the matter era and begins growing only after reheating.
Thus, the curvaton will have a chance to dominate (or nearly dominate)
the Universe only if
\begin{equation}
\Gamma<\Gamma_{\rm inf}\;.
\label{Gs}
\end{equation}
Using Eq.~(\ref{ratio4}), one obtains
\begin{equation}
\frac{\rho_\sigma}{\rho}
=\left(\frac{H_{\bar{m}}}{H}\right)^{1/2}
\left.\frac{\rho_\sigma}{\rho}\right|_m,
\label{ratiofinal}
\end{equation}
where 
\begin{equation}
H_{\bar{m}}\equiv\min\{m,H_m,\Gamma_{\rm inf}\}\;.
\label{Hbarm}
\end{equation}
The above suggests that, if the curvaton decays (or thermalizes) before 
domination, then, from Eq.~(\ref{r}), we have
\begin{equation}
r\sim \sqrt{H_{\bar{m}}/\Gamma}\;\times
\left.\frac{\rho_\sigma}{\rho}\right|_m,
\label{rG}
\end{equation}
If, on the other hand, the curvaton does manage to dominate, then 
Eq.~(\ref{ratiofinal}) gives
\begin{equation}
H_{\rm dom}\sim 
H_{\bar{m}}\left(\left.\frac{\rho_\sigma}{\rho}\right|_m\right)^2.
\label{Hdom}
\end{equation}

Hence, in the quasi-quadratic case, considering that 
\mbox{$H_{\bar{m}}=$ min\{$m,\Gamma_{\rm inf}$\}} and
in view of Eq.~(\ref{ratiomquad}), the above gives
\begin{eqnarray}
H_{\rm dom} & \sim & \min\{m,\Gamma_{\rm inf}\}
\left(\frac{\sigma_{\rm end}}{m_P}\right)^4
\left(\frac{m/\sqrt{c}}{H_{\rm end}}\right)^{2K_{\rm MD}}\times\nonumber\\
 & & \times
\min\!\left\{1,\frac{m/\sqrt{c}}{\Gamma_{\rm inf}}
\right\}^{2(K_{\rm RD}-K_{\rm MD})}.
\label{Hdomquad}
\end{eqnarray}

In the higher--order case there are three possibilities, as follows.
The first possibility is that all the higher--order oscillations occur
before reheating. Then, \mbox{$H_{\bar{m}}=\Gamma_{\rm inf}$} and
using Eqs.~(\ref{ratiomnonrenMD}) and (\ref{Hdom}) we obtain
\begin{equation}
H_{\rm dom}\sim\left(\frac{m}{H_{\rm osc}}\right)^2
\left(\frac{\sigma_{\rm end}}{m_P}\right)^4\Gamma_{\rm inf}\;.
\label{Hdomnonren1}
\end{equation}
The next possibility is that all the higher--order oscillations
occur after reheating, in which case \mbox{$H_{\bar{m}}=H_m$}. Then,
substituting Eqs.~(\ref{ratiomnonrenRD}) and (\ref{HmnonrenRD}) into 
Eq.~(\ref{Hdom}), we find
\begin{equation}
H_{\rm dom}\sim\left(\frac{m}{H_{\rm osc}}\right)
\left(\frac{\sigma_{\rm end}}{m_P}\right)^4m\,.
\label{Hdomnonren2}
\end{equation}
Finally, the last possibility is that
the higher--order oscillations begin before reheating but they continue 
afterwards into the radiation era. Again we have \mbox{$H_{\bar{m}}=H_m$}.
Substituting Eqs.~(\ref{ratiomnonrenMDRD}) and (\ref{HmnonrenMDRD}) into 
Eq.~(\ref{Hdom}), we get
\begin{equation}
H_{\rm dom}\sim\left(\frac{m}{H_{\rm osc}}\right)^2
\left(\frac{\sigma_{\rm end}}{m_P}\right)^4\Gamma_{\rm inf}\;,
\label{Hdomnonren3}
\end{equation}
which is identical to the first possibility (c.f. Eq.~(\ref{Hdomnonren1})).
Thus, in general, for the higher--order oscillations, we have found that
\begin{equation}
H_{\rm dom}\sim\left(\frac{m}{H_{\rm osc}}\right)^2
\left(\frac{\sigma_{\rm end}}{m_P}\right)^4
\min\{H_{\rm osc},\Gamma_{\rm inf}\}\,.
\label{Hdomnonren}
\end{equation}
From the above, we see that, remarkably, 
{\em in all cases $H_{\rm dom}$ is independent of $n$}. 

What if the quadratic term was dominant right from the start? This is indeed 
possible if
\begin{equation}
c_0<c<\left(\frac{m}{H_{\rm end}}\right)^2\ll 1\,.
\end{equation}
Then, \mbox{$(\rho_\sigma/\rho)_m=(\rho_\sigma/\rho)_{\rm osc}$}, which is 
given by Eq.~(\ref{ratioosc}). Thus, in this case, we find
\begin{equation}
H_{\rm dom}\sim\left(\frac{\sigma_{\rm end}}{m_P}\right)^4
\min\{m,\Gamma_{\rm inf}\}\,.
\end{equation}
Note that the above can be obtained by setting 
\mbox{$c\rightarrow(m/H_{\rm end})^2$}
in Eq.~(\ref{Hdomquad}) and \mbox{$H_{\rm osc}\rightarrow m$} in 
Eq.~(\ref{Hdomnonren}).

\subsection{\boldmath The value of $q$}\label{qsec}

In this section, we take a closer look to the value of $q$, which relates the 
fractional perturbation of the curvaton at horizon crossing and at curvaton 
decay as defined in Eq.~(\ref{qdef}). We have already seen that, in all cases 
(c.f. Eqs.~(\ref{dff1}) and (\ref{dff2})), 
\begin{equation}
\left.\frac{\delta\sigma}{\sigma}\right|_{\rm end}\simeq
\left.\frac{\delta\sigma}{\sigma}\right|_{\rm dec}\sim\zeta_\sigma\;.
\label{dff0}
\end{equation}
Below, however, we will study this relation in more detail. For simplicity,
we drop the bar from the oscillation amplitude of the field.

During inflation and after the fast-roll regime, the curvaton is overdamped.
As shown by Eq.~(\ref{frz0}), this means that the value of the curvaton 
perturbation is frozen. It is trivial to show, using Eq.~(\ref{SR}), that a 
very similar relation is true for the value of $\sigma_*$ itself. This is 
especially true if the curvaton has reached the quantum regime.
Thus, if $c$ does not change significantly at the end of 
inflation, it is safe to consider that
\begin{equation}
\left.\frac{\delta\sigma}{\sigma}\right|_*\approx
\left.\frac{\delta\sigma}{\sigma}\right|_{\rm end}.
\label{dff}
\end{equation}
Also, since we have shown that, before decaying, the curvaton needs to be
in the final quadratic oscillation regime, Eq.~(\ref{zf}) suggests that
\begin{equation}
\zeta_\sigma=\frac{2}{3}\left.\frac{\delta\sigma}{\sigma}\right|_{\rm dec}
=\frac{2}{3}q\left.\frac{\delta\sigma}{\sigma}\right|_*.
\label{zs}
\end{equation}

\subsubsection{The quasi-quadratic case}

In the quasi-quadratic case, Eqs.~(\ref{solu1}) and (\ref{solu2}) show clearly 
that 
\begin{equation}
\sigma(t)=\sigma_0\times f(t/t_0)\,,
\label{st}
\end{equation}
where $f$ is some function with \mbox{$f(1)=1$}, which is independent of 
$\sigma_0$. It is clear from the above that, if we consider a perturbed value 
of the field, we will have
\begin{equation}
\frac{\delta\sigma}{\sigma}(t)=
\frac{\delta\sigma_0}{\sigma_0}={\rm const.}\quad.
\label{dffquad}
\end{equation}
This is hardly surprising since, for potentials of the form 
\mbox{$V\propto\sigma^2$}, the equations of motion for the field and its 
perturbation (Eqs.~(\ref{KG}) and (\ref{dfKG}) respectively) are identical.
Now it is crucial to note that {\em the onset of the final quadratic 
oscillations occurs at the same time regardless of the value of the field}.
This is due to Eq.~(\ref{Hmqq}), which shows that $H_m$ is 
$\sigma$-independent. Thus, choosing \mbox{$t_0=t_{\rm end}$}, 
Eq.~(\ref{dffquad}) gives
\begin{equation}
\frac{\delta\sigma}{\sigma}(H_m)=
\left.\frac{\delta\sigma}{\sigma}\right|_{\rm end}.
\label{dffquad1}
\end{equation}
Eq.~(\ref{dffquad}) is obviously valid during the final quadratic 
oscillations as well, carrying the perturbations intact until the time of 
curvaton decay, which again is $\sigma$-independent. Thus, 
\begin{equation}
\left.\frac{\delta\sigma}{\sigma}\right|_{\rm dec}=
\frac{\delta\sigma}{\sigma}(H_m)\,.
\label{dffquad2}
\end{equation}
As explained in Sec.~\ref{qq1}, if \mbox{$c\ll 1$}, it is possible that the 
field briefly freezes again before the onset of the final quadratic 
oscillations, but this is not expected to affect the amplitude of the 
perturbation spectrum. All in all, in view also of Eq.~(\ref{dff}), we have 
found that, in the quasi-quadratic case
\begin{equation}
q=1\;,
\end{equation}
which means that 
\begin{equation}
\zeta_\sigma=\frac{2}{3}\left.\frac{\delta\sigma}{\sigma}\right|_*.
\label{zsquad}
\end{equation}

\subsubsection{The higher--order case}

The case of higher--order term domination is a bit more complicated.
Due to Eq.~(\ref{barf}), one easily sees that, in this case too, 
Eqs.~(\ref{st}) and (\ref{dffquad}) are valid. However, because 
\mbox{$m_{\rm eff}\sim\sigma^{n+2}/M^n$} and in contrast to the 
quasi-quadratic case, {\em the onset of both the higher-order and the final 
quadratic oscillations does depend on the value of} $\sigma$. 
Therefore, perturbing this value somewhat affects also the scaling of 
the field. Below we calculate what this means for $q$.

Suppose we consider two causally disconnected regions of the Universe, 
characterized by two different values for the curvaton field: $\sigma^{(1)}$
and $\sigma^{(2)}$. Suppose also that, in both these regions, the field lies 
in the higher-order regime. Then, in these two regions, the onset of the 
higher--order oscillations and the final quadratic oscillations will occur 
at different times. Without loss of generality, we assume that it is in 
region--(1) that the field begins its (Hubble damped)
sinusoidal oscillations first. Using
Eq.~(\ref{barf}), we obtain the following for the amplitudes of the field
at this moment:
\begin{eqnarray}
\sigma_{m;1}^{(1)} & = & \sigma^{(1)}_{\rm end}
\left(\frac{a^{(1)}_{\rm osc}}{a^{(1)}_m}\right)^{\frac{6}{n+6}}
,\nonumber\\
\sigma_{m;1}^{(2)} & = & \sigma^{(2)}_{\rm end}
\left(\frac{a^{(2)}_{\rm osc}}{a^{(1)}_m}\right)^{\frac{6}{n+6}}.
\label{sm1}
\end{eqnarray}
Considering the fact that \mbox{$a\propto H^{-2/3(1+w)}$} and also that 
\mbox{$H_{\rm osc}^2\sim m_{\rm eff}^2(\sigma_{\rm end})
\propto\sigma^{n+2}_{\rm end}$}, the above gives
\begin{equation}
\left.\frac{\sigma^{(2)}}{\sigma^{(1)}}\right|_{m;1}=
\left(\left.\frac{\sigma^{(2)}}{\sigma^{(1)}}\right|_{\rm end}
\right)^{1-\left(\frac{n_c+6}{n_c+2}\right)/\left(\frac{n+6}{n+2}\right)}.
\label{sfracm1}
\end{equation}

Now, after the onset of the final quadratic oscillations for $\sigma^{(1)}$,
its density decreases as \mbox{$\rho_\sigma\propto\sigma^2\propto a^{-3}$}.
On the other hand, $\sigma^{(2)}$ still oscillates in the higher--order regime,
until the final quadratic oscillations begin for $\sigma^{(2)}$ too. Therefore,
at this moment, we have
\begin{eqnarray}
\sigma_{m;2}^{(1)} & = & \sigma^{(1)}_{m;1}
\left(\frac{a^{(1)}_m}{a^{(2)}_m}\right)^{3/2}
,\nonumber\\
\sigma_{m;2}^{(2)} & = & \sigma^{(2)}_{m;1}
\left(\frac{a^{(1)}_m}{a^{(2)}_m}\right)^{\frac{6}{n+6}}.
\label{sm}
\end{eqnarray}

Let us assume now that, during the period of the higher--order oscillations, 
the equation of state of the Universe remains unmodified. Then we can use 
Eq.~(\ref{Hmnonren}) to relate $H_m$ with $H_{\rm osc}$. Using again that 
\mbox{$a\propto H^{-2/3(1+w)}$} and also that \mbox{$H_{\rm osc}^2\sim 
m_{\rm eff}^2(\sigma_{\rm end})\propto\sigma^{n+2}_{\rm end}$}, we find
\begin{equation}
\hspace{-.1cm}
\left.\frac{\sigma^{(2)}}{\sigma^{(1)}}\right|_{m;2}\hspace{-.3cm}=
\left.\frac{\sigma^{(2)}}{\sigma^{(1)}}\right|_{m;1}
\left(\left.\frac{\sigma^{(2)}}{\sigma^{(1)}}\right|_{\rm end}
\right)^{\frac{n+2}{4}
\left[1-\left(\frac{n_c+6}{n_c+2}\right)/\left(\frac{n+6}{n+2}\right)\right]}
\label{sfracm2}
\hspace{-.25cm}.
\end{equation}
Now, after $\sigma^{(2)}$ begins its final quadratic oscillations both
fields oscillate sinusoidally (with Hubble damping)
and, therefore, Eq.~(\ref{dffquad}) is valid.
Since the decay of the curvaton occurs independently of the value of $\sigma$,
we see that \mbox{$(\sigma^{(2)}/\sigma^{(1)})_{\rm dec}=
(\sigma^{(2)}/\sigma^{(1)})_{m;2}$}. Using this and
combining Eqs.~(\ref{sfracm1}) and (\ref{sfracm2}), we find that
\begin{equation}
\left.\frac{\sigma^{(2)}}{\sigma^{(1)}}\right|_{\rm dec}=
\left(\left.\frac{\sigma^{(2)}}{\sigma^{(1)}}\right|_{\rm end}
\right)^{\frac{n+6}{4}
\left[1-\left(\frac{n_c+6}{n_c+2}\right)/\left(\frac{n+6}{n+2}\right)\right]}.
\label{sfracm}
\end{equation}

Therefore, if we take that \mbox{$\sigma=\sigma^{(1)}$} and 
\mbox{$\sigma+\delta\sigma=\sigma^{(2)}$} and also 
\mbox{$\delta\sigma\ll\sigma$}, we obtain
\begin{equation}
\left.\frac{\delta\sigma}{\sigma}\right|_{\rm dec}\hspace{-.3cm}=
\frac{n+6}{4}
\left[1-\left(\frac{n_c+6}{n_c+2}\right)\Big/
\left(\frac{n+6}{n+2}\right)\right]
\left.\frac{\delta\sigma}{\sigma}\right|_{\rm end}\hspace{-.3cm}+\cdots\;,
\end{equation}
where the ellipsis denotes terms of higher power in 
$(\delta\sigma/\sigma)_{\rm end}$. Hence, in view also of Eq.~(\ref{dff}),
we have found that
\begin{equation}
q=\frac{1}{4}(n+6)
\left[1-\left(\frac{n_c+6}{n_c+2}\right)\Big/
\left(\frac{n+6}{n+2}\right)\right]\;,
\label{qnonren}
\end{equation}
which, for matter and radiation eras respectively, becomes
\begin{eqnarray}
q_{_{\rm MD}}=\frac{2-n}{4} & \quad{\rm and}\quad & 
q_{_{\rm RD}}=\frac{6-n}{8}\,.
\label{qMDRD}
\end{eqnarray}

From Eq.~(\ref{qnonren}), it is evident that, for \mbox{$n=n_c$}, we have
\mbox{$q=0$} and all the perturbation spectrum is erased. From 
Eq.~(\ref{sfracm}), it is clear that this is so to all orders in 
$(\delta\sigma/\sigma)_{\rm end}$. Hence, we see that, {\em the fact that the 
onset of the higher--order and the final quadratic oscillations is 
$\sigma$-dependent suppresses strongly the perturbation spectrum if 
\mbox{$n=n_c$}}. This is why the border case \mbox{$n=n_c$} for the attractor
constraint in Eq.~(\ref{attr}) is not acceptable, even though 
the field oscillates and does not converge towards the attractor solution.
For \mbox{$n<n_c$}, however, we see that \mbox{$q\simeq 1$} and so 
Eq.~(\ref{dff0}) is justified. 

In the case where the higher--order oscillations begin during matter domination
but end after reheating, working as above and using Eq.~(\ref{HmnonrenMDRD}) 
instead of Eq.~(\ref{Hmnonren}), one can show that
\begin{equation}
q_{_{\rm MD}}=\frac{(n+5)(2-n)}{3(n+6)}\,.
\end{equation}
Since, when unfreezing occurs during matter domination,
only the case of the quartic term is allowed by the attractor constraint in
Eq.~(\ref{attr}), we see that, in this case, \mbox{$q=1/2$} \{\mbox{$q=5/9$}\} 
if the curvaton decays before \{after\} reheating.

\subsubsection{Effective $q$ for significant change in $c$}\label{smallq}

Before concluding this section, we should point out the possibility that $c$
does change significantly at the end of inflation. This will not have an effect
if the field remains in the higher--order regime. However, if \mbox{$c\sim 1$}
after the end of inflation, this will probably relocate the field into the 
quasi-quadratic part of the potential. In this case, as discussed
also in Sec.~\ref{qq-c}, $(\delta\sigma/\sigma)_{\rm end}$ may be substantially
reduced. We will demonstrate this assuming that $\sigma_*$ has reached the 
quatum regime.

In view of Eq.~(\ref{fend-c}), one finds that Eq.~(\ref{dff}) becomes
\begin{equation}
\left.\frac{\delta\sigma}{\sigma}\right|_{\rm end}\sim
\left(\frac{\sigma_{\rm end}(c_*)}{\sigma_{\rm min}^{\rm end}(c)}\right)
\left.\frac{\delta\sigma}{\sigma}\right|_*,
\label{dffc}
\end{equation}
where \mbox{$c\sim 1$} corresponds to the value after the end of inflation.
Thus, since \mbox{$\sigma_{\rm end}(c_*)\sim\sigma_Q\ll
\sigma_{\rm min}^{\rm end}(c_*)<\sigma_{\rm min}^{\rm end}(c)$}, 
we see that {\em the effective value of $q$ may be substantially reduced if 
$c$ changes significantly at the end of inflation} in agreement with
Eq.~(\ref{zetaminus}).
Considering that \mbox{$\sigma_{\rm end}(c_*)\approx\sigma_*$} and also
that \mbox{$\delta\sigma_*\sim H_*$}, the above becomes
\begin{equation}
\left.\frac{\delta\sigma}{\sigma}\right|_{\rm end}\sim
\left(\frac{c_0}{c}\right)^{\frac{1}{n+2}}c_0\;.
\label{cc0}
\end{equation}
From Eq.~(\ref{df/fend}), we have 
\mbox{$(\delta\sigma/\sigma)_*\sim$ max\{$c_*,c_0$\}}, which, in view of the 
above, gives
\begin{equation}
q_{\rm eff}\sim\left(\frac{c_0}{c}\right)^{\frac{1}{n+2}}
\min\left\{1,\frac{c_0}{c_*}\right\}.
\label{qeff}
\end{equation}

\section{Thermal corrections}\label{Tcorr}

In this section, we introduce also the temperature correction to the potential,
as it appears in Eq.~(\ref{V0}). We will investigate the effect of this 
correction on the dynamics of the rolling and oscillating curvaton.
The mass of the curvaton is taken now to include all the contributions
\begin{equation}
V''\sim m^2\pm cH^2(t)+m_{\rm eff}^2+g^2T^2,
\label{mass}
\end{equation}
where, in this section, with $m_{\rm eff}$ we refer to the contribution due to 
the higher--order terms: \mbox{$m_{\rm eff}^2\sim\sigma^{n+2}/M^2$}.

During the matter era of the inflaton's oscillations, there exists a 
subdominant thermal bath due to the inflaton's decay products. One would 
naively expect that any thermal bath during a matter era is to be rapidly 
diluted away since radiation density redshifts faster than matter, according 
to Eq.~(\ref{raw}). However, due to the continuous decay of the inflaton field,
the thermal bath receives an ever-growing contribution to its density, so that 
the continuity equation, Eq.~(\ref{energy}), does not apply and, therefore, 
neither does Eq.~(\ref{raw}). The temperature of this thermal bath is
\cite{KT}
\begin{equation}
T\sim(m_P^2\Gamma_{\rm inf}H)^{1/4}.
\label{TMD}
\end{equation}
According to the above, the radiation density during the matter era scales as 
\mbox{$\rho_\gamma\propto T^4\propto H$}. Since this density is negligible
compared to the one of the oscillating inflaton field, the loss of energy 
from the latter is also negligible so that, for the inflaton, the continuity 
equation does apply. Now, from Eq.~(\ref{fried}), we have 
\mbox{$\rho\propto H^2$}, which means that there is a moment when the radiation
density becomes comparable to the one of the inflaton. This is the
moment of reheating and it occurs when 
\mbox{$H_{\rm reh}\sim\Gamma_{\rm inf}$}. In view of Eq.~(\ref{TMD}), the
reheat temperature is 
\begin{equation}
T_{\rm reh}\sim\sqrt{m_P\Gamma_{\rm inf}}\;.
\label{Treh}
\end{equation}
After reheating, the Universe becomes radiation dominated with 
\mbox{$\rho\sim T^4$}. In view of Eq.~(\ref{fried}), therefore, we can write 
for the temperature of the Universe
\begin{equation}
T\sim\Big(m_P^2H\min\{H,\Gamma_{\rm inf}\}\Big)^{1/4}
\propto H^{\frac{1+3w}{4}}.
\label{T}
\end{equation}

Due to its interaction with the thermal bath, the scalar condensate is in 
danger of evaporating, that is of becoming thermalized, if 
\mbox{$\Gamma_T>\Gamma_\sigma$}, i.e. if the decay rate of the field 
is smaller than the thermalization rate. If thermalization does occur, even
if, strictly speaking, the curvatons have not decayed yet into other particles,
they constitute a component of the thermal bath and their equation of state is 
that of radiation. Thus, after thermalization the density ratio of the 
curvaton to the radiation background remains constant. As a result, $r$ is 
decided not at the time of the curvaton decay but at the time of 
thermalization and this is why, so far, we have used  $\Gamma$
as given by Eq.~(\ref{G}) to determine 
the amplitude of the total curvature perturbation.

Let us estimate $\Gamma_T$. The scattering cross-section for scalar 
particles is \mbox{$\sigma\sim g^2/E_{\rm cm}^2$}, where 
\mbox{$E_{\rm cm}\sim\sqrt{Tm_\sigma}$} is the centre-of-mass energy. Thus, 
for the scattering rate \mbox{$\Gamma_{\rm sc}\sim\sigma T^3$}, we find 
\mbox{$\Gamma_{\rm sc}\sim g^2T^2/m_\sigma$}. When the particles thermalize,
\mbox{$m_\sigma\sim gT$} and so, for thermalization, we find
\begin{equation}
\Gamma_T\sim gT\,.
\label{GT}
\end{equation}
Suppose, now, that the temperature corrections were indeed dominating the 
effective mass of Eq.~(\ref{mass}). Then 
\mbox{$V''\sim g^2T^2\sim \Gamma_T^2$}.
As usual, the field is able to oscillate only when $H(t)$ drops enough so that
\mbox{$V''\sim H^2$}. However, in this case, Eq.~(\ref{GT}) suggests that,
when the field is about to begin oscillating, \mbox{$H\sim\Gamma_T$} and, 
therefore, the field thermalizes and the condensate evaporates. Hence, we see
that {\em the field thermalizes before engaging into oscillations, when the 
effective mass is dominated by the temperature corrections}.

As discussed in the previous section, the curvaton has a chance to dominate 
(or nearly dominate) the Universe only if it decays or thermalizes after the
onset of the final quadratic oscillations. Thus, in order to protect the 
curvaton from thermalizing too early, we have to demand that
\begin{equation}
gT_m<m\,,
\label{gTcons}
\end{equation}
where \mbox{$T_m\equiv T(H_m)$} is the temperature when $\sqrt{V''}$ becomes 
dominated by the soft mass $m$. Note that, for later times, \mbox{$gT<m$}
and the thermalization rate is suppressed. 

The above condition, Eq.~(\ref{gTcons}), suffices in order to avoid 
thermalization, only if, before \mbox{$H\sim H_m$}, the temperature corrections
have been always subdominant. This is ensured provided both $cH^2$ and
$m_{\rm eff}^2$ decrease with time faster than $(gT)^2$. This is indeed the 
case as we show below.

\begin{figure}[t]
\includegraphics[width=75mm,angle=0]{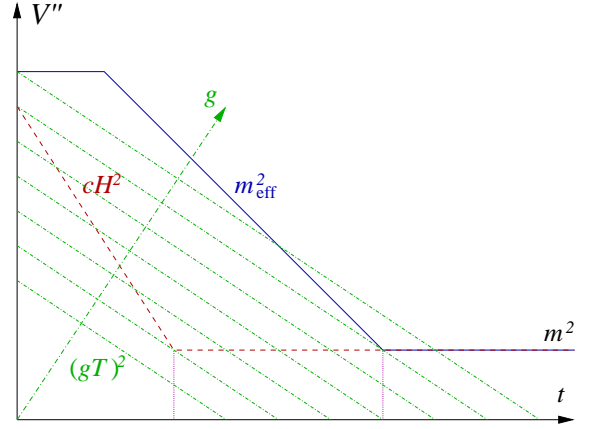}
\caption{\label{temp} 
Illustration (log-log plot) of the different scalings of the various
contributions to $V''$ assuming a given background barotropic parameter $w$. 
The solid line corresponds to the higher--order case, where 
\mbox{$V''\sim m_{\rm eff}^2\sim\sigma^{n+2}/M^n$}. The dashed line 
corresponds to the quasi-quadratic case, where \mbox{$V''\sim cH^2$}. The thin 
dot-dashed lines correspond to $(gT)^2$ for different values of $g$ growing as 
the thick dot-dashed arrow shows. It is depicted that 
the quasi-quadratic contribution to the effective mass decreases faster than 
the higher--order one and they both decrease faster than the temperature 
induced mass. From the figure, it is evident that, if $(gT)^2$ is subdominant 
when the soft mass--squared $m^2$ takes over, then this is enough to ensure 
that the temperature contribution to the effective mass is always subdominant,
both before and after the moment when $m^2$ takes over.}
\end{figure}

Consider first the quasi-quadratic case. Then, in view of Eq.~(\ref{T}), we 
find
\begin{equation}
\frac{cH^2}{(gT)^2}\propto H^{\frac{3}{2}(1-w)},
\label{mfracqq}
\end{equation}
which is indeed decreasing with the Universe expansion. Now, in the 
higher--order case, we have the following. Eqs.~(\ref{meff}) and (\ref{T}),
in view also of Eq.~(\ref{nc}), give
\begin{equation}
\frac{m_{\rm eff}^2}{(gT)^2}\propto 
H^{\frac{4}{1+w}\left(\frac{n+2}{n+6}\right)-\frac{1+3w}{2}}.
\label{mfracnonren}
\end{equation}
The exponent in the {\sc rhs} of the above is positive when 
\mbox{$n>-2\left(\frac{5-w}{7+5w}\right)$}, which is always true for 
\mbox{$n\geq 0$}. Thus, the effective mass $m_{\rm eff}$ also decreases faster 
than the $gT$ and, hence, {\em the condition in Eq.~(\ref{gTcons}) ensures 
that the temperature corrections to the effective mass are always subdominant}.
This is illustrated in Fig.~\ref{temp}.

The condition in Eq.~(\ref{gTcons}) can be translated into an upper bound on 
the coupling $g$ of the curvaton to the thermal bath. Indeed, from 
Eqs.~(\ref{T}) and (\ref{gTcons}), we find
\begin{equation}
g^2<\frac{m^2}{m_PH_m}
\max\left\{1,\frac{H_m}{\Gamma_{\rm inf}}\right\}^{1/2}.
\label{gbound}
\end{equation}
In the quasi-quadratic case, using Eq.~(\ref{Hmqq}), the above becomes
\begin{equation}
g^2<\sqrt{c}\;\frac{m}{m_P}
\max\left\{1,\frac{m/\sqrt{c}}{\Gamma_{\rm inf}}\right\}^{1/2}.
\label{gboundqq}
\end{equation}
Now, using Eqs.~(\ref{HmnonrenMD}), (\ref{HmnonrenRD}) and 
(\ref{HmnonrenMDRD}), in the higher--order case, we find
%
%
\begin{equation}
g^2<\frac{m}{m_P}
\left(\frac{H_{\rm osc}}{m}\right)^{\frac{2-n}{4(n+2)}}
\left(\frac{m}{\Gamma_{\rm inf}}\right)^{1/2},
\label{gboundnonrenMD}
\end{equation}
for 
\mbox{$\Gamma_{\rm inf}\leq H_m<H_{\rm osc}$},
\begin{equation}
g^2<\frac{m}{m_P}
\left(\frac{H_{\rm osc}}{m}\right)^{\frac{2}{3}\left(\frac{2-n}{n+2}\right)}
\left(\frac{m}{\Gamma_{\rm inf}}\right)^{-1/3},
\label{gboundnonrenMDRD}
\end{equation}
for \mbox{$H_m<\Gamma_{\rm inf}\leq H_{\rm osc}$}, and
\begin{equation}
g^2<\frac{m}{m_P}
\left(\frac{H_{\rm osc}}{m}\right)^{\frac{6-n}{3(n+2)}},
\label{gboundnonrenRD}
\end{equation}
for \mbox{$H_m<H_{\rm osc}<\Gamma_{\rm inf}$}\ .

As discussed above, the constraint in Eq.~(\ref{gTcons}) guarantees that the
temperature correction to the potential has negligible effect to the dynamics 
of the curvaton. In this case, $\Gamma_T$ becomes irrelevant and, therefore,
\mbox{$\Gamma=\Gamma_\sigma$}. For strong coupling $g$, we expect 
\mbox{$\Gamma_\sigma\sim g^2m$}. However, if $g$ is very small then the 
curvaton may have to decay gravitationally, in which case its decay will be 
suppressed by $m_P$. Thus, we can write
\begin{equation}
\Gamma_\sigma\sim m\;\max\left\{g^2,\left(\frac{m}{m_P}\right)^2\right\}.
\label{Gf}
\end{equation}

\section{A concrete example}\label{exam}

In this section, we will demonstrate our findings by employing them in a 
specific model, under our minimalistic assumption of a single inflationary 
period and a single curvaton field. By assuming certain quantities, such as 
the inflationary energy scale, the decay rate of the inflaton, the mass 
scale $M$ and so on, we aim to obtain bounds on the curvaton model and its 
parameters, such as $m$ or~$g$.

In particular, for our example, we choose the following values:
\begin{eqnarray}
M=m_P\;,
 & \qquad & \lambda_n\simeq 1
\end{eqnarray}
and also
\begin{eqnarray}
V_{\rm inf}^{1/4}\sim 10^{14}{\rm GeV}\,, & & 
H_*\sim 10^{11}{\rm GeV}\,,\nonumber\\
m_{\rm inf}\sim 10^{10}{\rm GeV}\,, & \qquad &
\Gamma_{\rm inf}\sim 10^{-6}{\rm GeV}\;.
\label{parameters}
\end{eqnarray}
In the above choice, we took into account the {\sc cobe} upper bound on the 
inflationary scale \mbox{$V_{\rm inf}^{1/4}<10^{16}$GeV} necessary to render
the inflaton's perturbations negligible \cite{dl}. Also, we considered that
the mass of the inflaton $m_{\rm inf}$ should be small enough for slow roll, 
\mbox{$m_{\rm inf}^2\ll H_*^2$}, but not too small since, for most inflation
models, \mbox{$\eta\sim(m_{\rm inf}/H_*)^2\sim 0.01$}. We chose $m_{\rm inf}$
to be of the so-called intermediate scale $\sim\sqrt{m_Pm_{\rm ew}}$, where
\mbox{$m_{\rm ew}\sim 1$ TeV} is the electroweak scale. Finally, we assumed 
that the inflaton is a gauge singlet and, therefore, decays only 
gravitationally so that \mbox{$\Gamma_{\rm inf}\sim m_{\rm inf}^3/m_P^2$}. The 
reheat temperature, for the above parameters, is 
\mbox{$T_{\rm reh}\sim 10^6$GeV} (c.f. Eq.~(\ref{Treh})), which is well above 
$T_{\rm BBN}$ but satisfies also comfortably the gravitino bound: 
\mbox{$T_{\rm reh}<10^9$GeV} \cite{note16}.

We assume that \mbox{$H_{\rm end}\approx H_*$} and also that
inflation lasts long enough for the curvaton to reach the quantum 
regime. Then Eq.~(\ref{fQ}) suggests
\begin{equation}
\hat{\sigma}_{\rm end}\sim\sigma_Q\sim H_*/\max\{c_*,c_0\}\;.
\label{send}
\end{equation}
This means that, in view of Eqs.~(\ref{qdef}), (\ref{r}),
(\ref{df/fend}), (\ref{dff1}) and (\ref{dff2}),
one finds
\begin{equation}
\zeta_0\sim r\,q\;\max\{c_*,c_0\}\;,
\label{z0}
\end{equation}
where \mbox{$\zeta_0=2\times 10^{-5}$}.
Using this, one can readily exclude the quartic case \mbox{$n=0$} regardless
of the choice of parameters in Eq.~(\ref{parameters}). Indeed,
reinstating the $\lambda_n$ in Eq.~(\ref{c0}), one finds
\begin{equation}
c_0\sim\frac12(\lambda_0/3)^{1/3}\sim 0.1\;.
\end{equation}
This means that, since \mbox{$c_*\ll 1$}, we have \mbox{$c_*<c_0$}. Thus,
using Eq.~(\ref{z0}) and employing also the {\sc wmap} constraint in 
Eq.~(\ref{wmap}), it is easy to find that we require
\begin{equation}
q\leq 10^{-2},
\end{equation}
which is possible to realize only if \mbox{$q=q_{\rm eff}$} since, otherwise,
\mbox{$q\sim 1$}. Thus, we need to have a change of sign  for the 
quasi-quadratic term after the end of inflation. Now, using Eq.~(\ref{qeff})
in this case, we obtain
\begin{equation}
q_{\rm eff}\sim\sqrt{c_0/c}\;,
\end{equation}
which can never be as small as $10^{-2}$. Hence, the quartic case is excluded,
unless $\lambda_0$ is extremely small or inflation ends well before the 
curvaton reaches the quantum regime.
The remaining two possibilities are \mbox{$n=2,4$} (we assume that symmetries 
in the potential allow only even powers). In the following, we choose 
\mbox{$n=4$}, because it offers the largest parameter space. Using this,
Eqs.~(\ref{c0}) and (\ref{parameters}) give
\begin{equation}
c_0\sim 10^{-4}.
\label{c0ex}
\end{equation}

We consider two cases. In the first case, we assume that the (approximate) 
symmetry which keeps $c$ small during inflation persists after inflation ends, 
so that \mbox{$c=c_*\ll 1$}. In the second case, we assume that 
\mbox{$c\simeq 1$} after the end of inflation, considering both the $\pm c$
cases. In all cases, we assume that $c$ does not change significantly at the 
transition between the matter and radiation dominated epochs.

\subsection{\boldmath Case $c=c_*\ll 1$}

In this case, there is no change of sign for the quasi-quadratic term so
we will drop the hat on $\sigma$, since, when both the constraints of
Eqs.~(\ref{attr}) and (\ref{nmin}) are satisfied, the curvaton's evolution
does not depend on the sign in front of $c$. 

Suppose, at first, that \mbox{$c\leq c_0$}. Then the field lies into the
higher--order part of the potential, where the attractor constraint in
Eq.~(\ref{attr}) is applicable. For \mbox{$n>0$}, we see that the curvaton
must unfreeze after reheating, that is
\begin{equation}
\Gamma_{\rm inf}>H_{\rm osc}\;.
\label{GHosc}
\end{equation}
Now, in view of Eq.~(\ref{Hosc}), we find
\begin{equation}
H_{\rm osc}\sim\sqrt{c_0}\,H_*\sim 10^9{\rm GeV}\;,
\label{Hoscex}
\end{equation}
where we used Eqs.~(\ref{parameters}) and (\ref{c0ex}).
From Eqs.~(\ref{parameters}) and (\ref{Hoscex}),
it is clear that Eq.~(\ref{GHosc}) is not satisfied and, therefore, the 
possibility of higher--order evolution (i.e. \mbox{$c\leq c_0$}) is excluded.

Let us assume now that \mbox{$c_0<c\ll 1$}, which results in
quasi-quadratic evolution. Since \mbox{$q=1$}, in this case, Eq.~(\ref{z0}) 
suggests that
\begin{equation}
r\sim 10^{-5}/c\;.
\end{equation} 
Using the {\sc wmap} constraint
of Eq.~(\ref{wmap}), we find that \mbox{$c\lsim 10^{-3}$}. However,
from Eq.~(\ref{c0ex}), we have \mbox{$c>c_0\sim 10^{-4}$}, which allows only 
one possibility
\begin{equation}
c\sim 10^{-3}
\label{cex}
\end{equation}
and, therefore,
\begin{equation}
r\sim 10^{-2}\;,
\end{equation}
which marginally satisfies Eq.~(\ref{wmap}).

Now, since \mbox{$c\ll 1$}, Eqs.~(\ref{KMD}) and (\ref{KRD}) suggest that
\mbox{$K_{\rm MD}\simeq K_{\rm RD}\simeq 0$} (c.f. Eq.~(\ref{Kc0})). 
Then, using Eq.~(\ref{ratiomquad}) in Eq.~(\ref{rG}), we obtain
\begin{equation}
r\sim\left(\frac{\sigma_{\rm end}}{m_P}\right)^2
\left(\frac{H_{\bar{m}}}{\Gamma_\sigma}\right)^{1/2}.
\label{rquad1}
\end{equation}
Eq.~(\ref{send}) gives \mbox{$\sigma_{\rm end}\sim 10^{14}$GeV}.
Substituting this into Eq.~(\ref{rquad1}) and using also 
Eq.~(\ref{parameters}), we find the condition
\begin{equation}
\min\{m, \Gamma_{\rm inf}\}\sim 10^{12}\Gamma_\sigma\,,
\label{condition}
\end{equation}
where we have also employed Eqs.~(\ref{Hmqq}) and (\ref{Hbarm}). To study this,
we need to use Eq.~(\ref{Gf}). Thus, we consider the following two cases:

\begin{itemize}
\item
{\bf\boldmath Small $g$:}
This case corresponds to
\begin{eqnarray}
g<\frac{m}{m_P} & \quad\Rightarrow\quad & \Gamma_\sigma\sim\frac{m^3}{m_P}\;,
\label{gsmall}
\end{eqnarray}
which means that the curvaton decays gravitationally. Using the above, the
{\sc bbn} constraint in Eq.~(\ref{bbn}) results in the bound
\begin{equation}
m> 10^4{\rm GeV}\;.
\label{bbnbound}
\end{equation}
Hence, in view also of Eq.~(\ref{parameters}), we see that 
\mbox{$m\gg\Gamma_{\rm inf}$}. Consequently, Eq.~(\ref{condition}) gives
\begin{equation}
m\sim 10^6{\rm GeV}\;,
\label{m1}
\end{equation}
which satisfies the bound of Eq.~(\ref{bbnbound}). Inserting the 
above into Eq.~(\ref{gsmall}), we obtain
\begin{equation}
g<10^{-12}\;,
\end{equation}
which, in view of Eqs.~(\ref{cex}) and (\ref{m1}),
can be easily shown to satisfy the bound of Eq.~(\ref{gboundqq}).

\item
{\bf\boldmath Large $g$:}
This case corresponds to
\begin{eqnarray}
g\geq\frac{m}{m_P} & \quad\Rightarrow\quad & \Gamma_\sigma\sim g^2m\;.
\label{glarge}
\end{eqnarray}
Using this, Eq.~(\ref{condition}) is recast as
\begin{equation}
\min\left\{1,\frac{\Gamma_{\rm inf}}{m}\right\}\sim 10^{12}g^2\;.
\label{cond}
\end{equation}
Taking \mbox{$m<\Gamma_{\rm inf}$}, one finds \mbox{$g\sim 10^{-6}$},
which can be easily shown, however, to violate the constraint of 
Eq.~(\ref{gboundqq}). Thus, we have
\begin{equation}
m\geq\Gamma_{\rm inf}\sim 10^{-6}{\rm GeV}\;.
\end{equation}
In view of the above, Eq.~(\ref{cond}) is now recast as
\begin{equation}
g\sim 10^{-9}(m/{\rm GeV})^{-1/2}.
\label{cond1}
\end{equation}
Using this and Eq.~(\ref{glarge}), one can show that the {\sc bbn} constraint 
in Eq.~(\ref{bbn}) is trivially satisfied. Further, Eq.~(\ref{glarge}) demands
that
\begin{equation}
m\leq 10^6{\rm GeV}\;.
\label{m2}
\end{equation}
Finally, employing the constraint of Eq.~(\ref{gboundqq}), we find
the bound
\begin{equation}
m\geq 0.1\;{\rm GeV}\;.
\label{m3}
\end{equation}

\end{itemize}

\begin{figure}[t]
\includegraphics[width=75mm,angle=0]{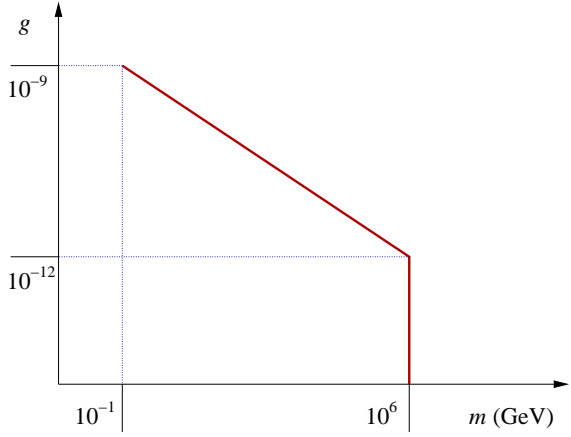}
\caption{\label{param} 
Illustration (log-log plot) of the parameter space (solid line) for a 
successful curvaton in the case when \mbox{$c=c_*\sim 10^{-3}$}, with model 
parameters given by Eq.~(\ref{parameters}). The upper bound 
\mbox{$g\leq 10^{-9}$} is determined by the requirement of 
Eq.~(\ref{gboundqq}), which ensures that $(gT)^2$ never dominates the 
effective mass. For \mbox{$g<10^{-12}$}, the curvaton decays predominantly 
through gravitational couplings.}
\end{figure}

Summing up, in this case, we have ended up with the following parameter space.
For large $g$, we have found
\begin{eqnarray}
10^{-1}{\rm GeV} \leq & m  & \leq 10^6{\rm GeV}\,,\nonumber\\
10^{-9} \geq & g & \geq 10^{-12},
\label{param1}
\end{eqnarray}
where the relation between $m$ and $g$ is given in Eq.~(\ref{cond1}). For 
small $g$, we have found
\begin{eqnarray}
 m & \sim & 10^6{\rm GeV}\,,\nonumber\\
 g & < & 10^{-12}.
\label{param2}
\end{eqnarray}
The above parameter space is shown in Fig.~\ref{param}. Note that, for all the 
parameter space, the masslessness requirement \mbox{$m\ll H_*\sim 10^{11}$GeV} 
is well satisfied.

\subsection{\boldmath Case $c\simeq 1\gg c_*$}

In this case, Eqs.~(\ref{z0}) and (\ref{c0ex}) yield
\begin{equation}
rq\leq 0.1\;.
\label{rq}
\end{equation}

The results depend on the sign of $c$. We treat the two cases separately.

\subsubsection{The case of $+c$}

In this case, \mbox{$q\sim 1$} and, therefore, Eq.~(\ref{rq}) suggests that
\begin{equation}
10^{-2}\leq r\leq 10^{-1},
\label{rrange}
\end{equation}
where we also took into account the {\sc wmap} constraint in Eq.~(\ref{wmap}).
The above and Eq.~(\ref{z0}) suggest that
\begin{equation}
1\leq\gamma\leq 10\;,
\label{grange}
\end{equation}
where we defined
\begin{equation}
\gamma\equiv\max\{1,c_*/c_0\}\;.
\label{gamma}
\end{equation}
Using this and Eqs.~(\ref{parameters}) and (\ref{send}), we obtain
\begin{equation}
\sigma_{\rm end}\sim\gamma^{-1}10^{15}{\rm GeV}\;.
\label{sendex}
\end{equation}

Now, when \mbox{$c\simeq 1$}, it can be shown that, after inflation,
the field finds itself in the quasi-quadratic regime. 
Eq.~(\ref{rrange}) shows that the curvaton has to decay before it dominates.
For \mbox{$c\geq c_{\rm x}$}, Eqs.~(\ref{KMD}) and (\ref{KRD}) show that
\mbox{$K_{\rm MD}=1$} and \mbox{$K_{\rm RD}=1/2$}. Substituting this in 
Eq.~(\ref{ratiomquad}), we find, from Eq.~(\ref{rG}), that
\begin{equation}
r\sim\left(\frac{\sigma_{\rm end}}{m_P}\right)^2
\left(\frac{m}{H_*}\right)
\left(\frac{\Gamma_{\rm inf}}{\Gamma_\sigma}\right)^{1/2}.
\label{rquad2}
\end{equation}
By comparing the above with Eq.~(\ref{rquad1}), we see that $r$ is now smaller
by a factor $(m/H_*)$. This is because, in contrast to the case 
\mbox{$c\ll 1$} when the curvaton's density ratio $\rho_\sigma/\rho$ remains 
almost constant after the end of inflation, in the case \mbox{$c\simeq 1$} 
this ratio is substantially reduced until the onset of the final quadratic 
oscillations. This makes things much more difficult and, as a result, the 
parameter space is diminished.

Now, employing Eqs.~(\ref{parameters}) and (\ref{sendex}), one finds from
Eq.~(\ref{rquad2}) the condition
\begin{equation}
m^2\sim \Gamma_\sigma\times \gamma^2 10^{38}{\rm GeV}\;.
\label{cond2}
\end{equation}
In view of the {\sc bbn} constraint in Eq.~(\ref{bbn}), the above gives the 
bound
\begin{equation}
m>\gamma 10^7{\rm GeV}\,.
\label{mbound}
\end{equation}

Now suppose that $g$ is small enough for Eq.~(\ref{gsmall}) to be valid.
Then, Eq.~(\ref{cond2}) gives \mbox{$m\sim\gamma^{-2}10^{-2}$GeV}, which 
strongly violates the bound in Eq.~(\ref{mbound}) for the whole range of 
$\gamma$. Thus, we have to consider the case when $g$ is large and
Eq.~(\ref{glarge}) is valid. Using this, Eq.~(\ref{cond2}) can be recast as
\begin{equation}
g\sim \gamma^{-1}10^{-19}(m/{\rm GeV})^{1/2}\;.
\label{cond3}
\end{equation}
In view of the above, the lower bound on $g$ from Eq.~(\ref{glarge}) gives 
\mbox{$m\leq\gamma^{-2}10^{-2}$GeV} which again violates Eq.~(\ref{mbound})
for the whole range of $\gamma$. Hence, in this case there is no parameter 
space for a successful curvaton.

\subsubsection{The case of $-c$}

In this case, \mbox{$q=q_{\rm eff}$} as given in Eq.~(\ref{qeff}). Using this 
and Eq.~(\ref{z0}), one, generically, obtains
\begin{equation}
\zeta_0\sim rc_0\left(\frac{c_0}{c}\right)^{\frac{1}{n+2}}.
\label{z0ex}
\end{equation}
In view of Eq.~(\ref{c0ex}) in our example, the above gives
\begin{equation}
r\sim 1
\end{equation}
and, therefore, it is possible for the curvaton to dominate the Universe before
its decay. Using Eqs.~(\ref{qeff}), (\ref{send}) and (\ref{z0ex}), it is easy 
to show that, generically,
\begin{eqnarray}
\sigma_{\rm min}(c,H_*)\gg\hat{\sigma}_{\rm end}
& \Leftrightarrow & q_{\rm eff} < 1\;.
\end{eqnarray}
Thus, if \mbox{$q_{\rm eff}<1$}, compared to the minimum after the end of 
inflation, the field appears to be at the origin. Therefore, the initial
amplitude of the quasi-quadratic oscillations is
\begin{equation}
\bar{\sigma}_{\rm end}\sim\sigma_{\rm min}(c,H_*)\;.
\end{equation}
Using this in our example, Eqs.~(\ref{fmin}) and (\ref{parameters}) suggest
\begin{equation}
\bar{\sigma}_{\rm end}\sim 10^{15}{\rm GeV}\;.
\label{sendex2}
\end{equation}

As before, for \mbox{$c\geq c_{\rm x}$}, Eqs.~(\ref{KMD}) and (\ref{KRD}) 
show that \mbox{$K_{\rm MD}=1$} and \mbox{$K_{\rm RD}=1/2$}. Substituting this 
in Eq.~(\ref{Hdomquad}), we obtain
\begin{equation}
H_{\rm dom}\sim
\left(\frac{\bar{\sigma}_{\rm end}}{m_P}\right)^4
\left(\frac{m}{H_*}\right)^2\Gamma_{\rm inf}\;.
\label{Hdomex}
\end{equation}
Inserting Eqs.~(\ref{parameters}) and (\ref{sendex2}) in the above, we find
\begin{equation}
H_{\rm dom}\sim 10^{-40}(m/{\rm GeV})^2{\rm GeV}\;.
\label{condex}
\end{equation}
Since the curvaton decays after domination, we require that
\begin{equation}
\Gamma_\sigma\leq H_{\rm dom}\;.
\label{GHdom}
\end{equation}
Combining Eqs.~(\ref{condex}) and (\ref{GHdom}) with the {\sc bbn} constraint
in Eq.~(\ref{bbn}), we find the bound
\begin{equation}
m>10^8{\rm GeV}\;.
\label{mboundex}
\end{equation}

Now suppose that $g$ is small so that Eq.~(\ref{gsmall}) is valid.
Then, Eqs.~(\ref{condex}) and (\ref{GHdom}) give 
\mbox{$m\leq 10^{-4}$GeV}, which strongly violates the bound in 
Eq.~(\ref{mboundex}). Thus, we have to consider the case when $g$ is large
and Eq.~(\ref{glarge}) is valid instead.
Using this, Eqs.~(\ref{condex}) and (\ref{GHdom}) result in
\begin{equation}
g\leq 10^{-20}(m/{\rm GeV})^{1/2}.
\label{bound1}
\end{equation}
It can be easily shown that, when the above is true the constraint of 
Eq.~(\ref{gboundqq}) is also satisfied.

Now the lower bound on $g$ coming from Eq.~(\ref{glarge}) is
\begin{equation}
g\geq 10^{-18}(m/{\rm GeV})\,.
\label{bound2}
\end{equation}
One can be easily check that this bound, in view of Eq.~(\ref{glarge}),
suffices to satisfy also the {\sc bbn} constraint in Eq.~(\ref{bbn}).

The available parameter space is spanned by Eqs.~(\ref{bound1}) and 
(\ref{bound2}) and it is a surface on the $m-g$ plane, in contrast to being a 
line, as shown in Fig.~\ref{param} for the previous case. This is because, 
when the curvaton dominates, the necessary condition is the inequality in
Eq.~(\ref{GHdom}), whereas if it decays before domination the necessary 
condition is the equality in Eq.~(\ref{rG}) (i.e. $r$ has to have exactly 
the right value). 

Unfortunately, from Eqs.~(\ref{bound1}) and (\ref{bound2}), we find that the 
parameter space exists only if \mbox{$m<10^{-4}$GeV}, which strongly violates 
the bound in Eq.~(\ref{mboundex}). Therefore, we have shown that there is no
parameter space for a successful curvaton in the case when \mbox{$c\simeq 1$}
regardless of the sign of $c$. This is due to the fact that the
quasi-quadratic oscillations drastically reduce the energy density of the 
field. This reduction is impossible to be counteracted during the 
final quadratic oscillations.

\section{Discussion and conclusions}\label{concl}

In this paper, we have investigated and analysed the dynamics of the curvaton 
field. Apart from its soft mass $m$, we have assumed that the field receives 
a contribution to its effective mass from supergravity 
corrections. This contribution is determined by the Hubble parameter and, 
hence, the potential acquires what we called the quasi-quadratic 
term. Another contribution to the effective mass may arise due to the 
presence of a quartic term or of non-renormalizable terms. 
Finally, we also considered possible thermal corrections to the potential,
which arise due to the coupling of the curvaton to the thermal bath 
present after the end of inflation.

Firstly, we discussed the behaviour of the field during inflation. We explained
that the field is expected to engage, initially, in fast-roll, which soon, 
however, is terminated to be followed by slow-roll evolution. Eventually, slow 
roll is also halted due to the action of field perturbations generated by 
quantum fluctuations. These perturbations effectively stabilize the coherent 
motion of the field for the remaining part of inflation.

After inflation, we have studied the evolution of the curvaton's energy 
density in the cases when the effective mass is dominated by the 
quasi-quadratic term or a higher--order one.
In the case of the quasi-quadratic domination, we have shown that there are
two types of evolution depending on how suppressed is the effective mass.
For effective mass of order the Hubble parameter, we found that the field 
begins to oscillate, whereas, for suppressed mass, we have found a scaling 
solution, according to which the field rolls gently towards the minimum of its 
potential. Both solutions preserve the amplitude of the perturbation spectrum 
of the curvaton, obtained during inflation. 

In the case of the higher--order term domination, we have shown again that the 
field, even though frozen initially, will, eventually, begin to oscillate 
(preserving also the amplitude of the perturbation spectrum) provided the 
order of the higher--order term is not too high. This is because, in the 
opposite case, after unfreezing, the field was found to follow an attractor 
solution, which loses all memory of initial conditions. Consequently, were the
field to follow such an attractor, all the superhorizon spectrum of its 
perturbations would have been erased and the field could not act as a curvaton.

A similar situation was also found in the quasi-quadratic case, when the 
effective mass was negative after the end of inflation. In this case, 
if the order of the higher--order term (which decides the temporal 
position of the minimum of the potential) is too high, instead of rolling, 
the field was found to follow a particular solution and remain at fixed,
constant ratio with respect to the temporal minimum, losing thereby all memory 
of initial conditions. The constraints obtained on the order of the principle
higher--order term in both the above cases were comparable and their
implementation was decisive for the evolution of the field. 

Generically, we have found the following behaviour for the curvaton after the 
end of inflation. Both in the quasi-quadratic case and the higher--order
case, the energy density of the field was found to be decreasing faster than 
the background density. Consequently, curvaton domination (or near 
domination) is possible only after the effective mass of the field becomes 
dominated by the soft mass. In all cases, when the attractor was avoided, we 
have found that the damping on the amplitude of the perturbation spectrum is 
negligible. Finally, we have shown that the temperature corrections should not 
be allowed to dominate the effective mass of the field because if they do 
the field thermalizes before having a chance to dominate (or nearly 
dominate). This requirement introduces a stringent constraint on the coupling 
of the curvaton to the thermal bath.

In all cases, we have calculated the density ratio $r$ of the curvaton to the 
background density of the Universe at the time of the curvaton's decay.
If the curvaton is to decay before dominating, this density ratio is 
necessary to determine the 
total curvature perturbation imposed by the curvaton field onto the Universe. 
We have also calculated when the curvaton dominates, if it does indeed so, in
all cases considered. Furthermore, we calculated in detail the factor $q$ 
relating the curvaton perturbation at horizon crossing with the one at its 
decay. Finally, we have employed our findings in a concrete model 
realization, for which we have shown that there is hardly any parameter space 
available for a successful curvaton if \mbox{$c\sim 1$} after the end of 
inflation, regardless of the sign in front of $c$. This is due to the fact
that, for large values of $c$, the density fraction of the curvaton is
drastically reduced by either the quasi-quadratic oscillations or the scaling 
solution. As explained also in Sec.~\ref{qq1}, we expect this problem to be 
generic and, if so, this would favour models where $c$ after inflation remains 
strongly suppressed, such as the {\sc pngb} curvaton models discussed in 
Ref.~\cite{dlnr}.

We should mention here that, due to the fact that our attempt to describing 
the curvaton evolution was aimed at the bulk of the parameter space, there are
possibly special situations, not studied here, where the evolution may be 
significantly modified.
For example, if the quasi-quadratic term is dominant but negative after the 
end of inflation and the field is undergoing oscillations, there will be a 
moment, just before the soft mass taking over, when the quadratic and 
quasi-quadratic terms cancel each-other. For a brief period, therefore, the 
field will be oscillating into a potential of higher order and, hence, will 
have an equation of state with a different (larger) effective barotropic 
parameter as shown by Eq.~(\ref{wf}). The outcome of all this will be a 
``glitch'' to the scaling of the density ratio that may affect both the time 
of curvaton domination and/or the total curvature perturbation if the curvaton 
decays before domination. The importance of such an effect is possible to be
studied only numerically.

Another example, in the case of negative quasi-quadratic mass, is the effect
of the local maximum at the origin, which may be felt if the initial condition
of the field is \mbox{$\sigma_{\rm end}\sim\sigma_{\rm min}$}, which is exactly
what we expect in this case, if the sign of $c$ changes at the end of 
inflation. Indeed, if the field engages into oscillations, there is a chance 
that, at some point, the oscillation amplitude marginally sends the field on 
top of the local maximum. The effect of such an event is the possible 
tachyonic amplification of the perturbation spectrum. In an analogous 
situation, this tachyonic amplification (when the field, as it oscillates, is 
temporarily stabilized near a local maximum of the potential) was thoroughly 
investigated by us in an earlier paper \cite{dllr1}, where it was found that 
the amplification of the perturbations could be substantial. One could even 
imagine that such an amplification mechanism may counteract the 
suppression of the amplitude of the perturbation spectrum due to 
attractor evolution.

Furthermore, we can imagine a number of modifications to our considerations
arising due to a Universe history more complicated than the one considered 
here. For example, one can insert a brief period of thermal inflation
before or after curvaton domination. Another example would be to consider a 
period of `kination', for which the background density has a stiff equation of 
state \mbox{$w\simeq 1$}, which arises naturally in models of quintessential 
inflation \cite{kostas,giov} or non-oscillatory inflationary models \cite{fl}. 
Indeed, the curvaton is a convenient mechanism to increase the effective 
reheating efficiency of such models \cite{lu}. 
We have not considered such possible complications in this paper 
mainly because there is an infinity of them. However, we tried to express our 
results in as much a model independent way as possible, so that they will be 
easy to implement in particular, more complicated scenarios of the Universe 
history.

All in all, we have investigated the evolution of the curvaton field during 
and after inflation. We have shown that a successful curvaton either 
oscillates or scales down to its minimum but always preserves the amplitude of 
the spectrum of its perturbations. We also showed that, in order to dominate 
(or nearly dominate), the curvaton has to decay after its effective mass 
becomes dominated by the soft mass term and the oscillation becomes 
(Hubble damped) sinusoidal. Furthermore, to 
avoid premature thermalization, the temperature corrections to the curvaton's 
potential should remain negligible throughout all its evolution. Finally, 
there is a part of the parameter space which results in destructive attractor 
evolution, which erases the perturbation spectrum and should be avoided.

\medskip

{\bf Acknowledgments:} We would like to thank D.~Wands for discussions.
This work was supported by the E.U. 5th-framework network programs
`Supersymmetry and the Early Universe': HPRN-CT-2000-00152 and
`Physics Across the Present Energy Frontier': HPRN-CT-2000-00148.

\def\ijmp#1#2#3{{Int. Jour. Mod. Phys.}
{\bf #1},~#3~(#2)}
\def\plb#1#2#3{{Phys. Lett. B }{\bf #1},~#3~(#2)}
\def\zpc#1#2#3{{Z. Phys. C }{\bf #1},~#3~(#2)}
\def\prl#1#2#3{{Phys. Rev. Lett.}
{\bf #1},~#3~(#2)}
\def\rmp#1#2#3{{Rev. Mod. Phys.}
{\bf #1},~#3~(#2)}
\def\prep#1#2#3{{Phys. Rep. }{\bf #1},~#3~(#2)}
\def\prd#1#2#3{{Phys. Rev. D }{\bf #1},~#3~(#2)}
\def\npb#1#2#3{{Nucl. Phys. }{\bf B#1},~#3~(#2)}
\def\npps#1#2#3{{Nucl. Phys. B (Proc. Sup.)}
{\bf #1},~#3~(#2)}
\def\mpl#1#2#3{{Mod. Phys. Lett.}
{\bf #1},~#3~(#2)}
\def\arnps#1#2#3{{Annu. Rev. Nucl. Part. Sci.}
{\bf #1},~#3~(#2)}
\def\sjnp#1#2#3{{Sov. J. Nucl. Phys.}
{\bf #1},~#3~(#2)}
\def\jetp#1#2#3{{JETP Lett. }{\bf #1},~#3~(#2)}
\def\app#1#2#3{{Acta Phys. Polon.}
{\bf #1},~#3~(#2)}
\def\rnc#1#2#3{{Riv. Nuovo Cim.}
{\bf #1},~#3~(#2)}
\def\ap#1#2#3{{Ann. Phys. }{\bf #1},~#3~(#2)}
\def\ptp#1#2#3{{Prog. Theor. Phys.}
{\bf #1},~#3~(#2)}
\def\apjl#1#2#3{{Astrophys. J. Lett.}
{\bf #1},~#3~(#2)}
\def\n#1#2#3{{Nature }{\bf #1},~#3~(#2)}
\def\apj#1#2#3{{Astrophys. J.}
{\bf #1},~#3~(#2)}
\def\anj#1#2#3{{Astron. J. }{\bf #1},~#3~(#2)}
\def\mnras#1#2#3{{MNRAS }{\bf #1},~#3~(#2)}
\def\grg#1#2#3{{Gen. Rel. Grav.}
{\bf #1},~#3~(#2)}
\def\s#1#2#3{{Science }{\bf #1},~#3~(#2)}
\def\baas#1#2#3{{Bull. Am. Astron. Soc.}
{\bf #1},~#3~(#2)}
\def\ibid#1#2#3{{\it ibid. }{\bf #1},~#3~(#2)}
\def\cpc#1#2#3{{Comput. Phys. Commun.}
{\bf #1},~#3~(#2)}
\def\astp#1#2#3{{Astropart. Phys.}
{\bf #1},~#3~(#2)}
\def\epjc#1#2#3{{Eur. Phys. J. C}
{\bf #1},~#3~(#2)}
\def\nima#1#2#3{{Nucl. Instrum. Meth. A}
{\bf #1},~#3~(#2)}
\def\jhep#1#2#3{{J. High Energy Phys.}
{\bf #1},~#3~(#2)}


\begin{thebibliography}{???}

\bibitem{wmapspergel}
D.~N.~Spergel {\it et al.},
astro-ph/0302209.

\bibitem{wmappeiris}
H.~V.~Peiris {\it et al.},
astro-ph/0302225.

\bibitem{wmapng}
E.~Komatsu {\it et al.},
astro-ph/0302223.

\bibitem{wmap}
{\sf http://map.gsfc.nasa.gov/}

\bibitem{note1}
In special cases, such as the one studied in 
Ref.~\cite{dllr1}, two or more scalar fields might be involved. As an 
alternative to the inflation hypothesis, a `pre-big-bang' 
\cite{prebigbang,BGGMV,LWC} or `ekpyrotic' \cite{ekpyrotic1,ekpyrotic2,pyro}
era of collapse has been proposed, but there is so far no accepted 
theory of a  bounce and therefore no firm prediction from collapsing 
cosmologies. In particular, there is so far no accepted string-theoretic 
description of a bounce \cite{lms}.

\bibitem{dllr1}
K.~Dimopoulos, G.~Lazarides, D.~Lyth and R.~Ruiz de Austri,
J. High Energy Phys. {\bf 05}, 057 (2003).

\bibitem{prebigbang}
G.~Veneziano,
hep-th/0002094.

\bibitem{BGGMV}
R.~Brustein, M.~Gasperini, M.~Giovannini, V.~F.~Mukhanov and
G.~Veneziano,
Phys. Rev. D {\bf 51}, 6744 (1995).

\bibitem{LWC}
J.~E.~Lidsey, D.~Wands and E.~J.~Copeland,
Phys.\ Rept.\  {\bf 337}, 343 (2000).

\bibitem{ekpyrotic1}
J.~Khoury, B.~A.~Ovrut, P.~J.~Steinhardt and N.~Turok,
Phys.\ Rev.\ D {\bf 64}, 123522 (2001);
D.~H.~Lyth,
Phys.\ Lett.\ B {\bf 524}, 1 (2002).

\bibitem{ekpyrotic2}
J.~Khoury, B.~A.~Ovrut, P.~J.~Steinhardt and N.~Turok,
Phys.\ Rev.\ D {\bf 66}, 046005 (2002);
D.~H.~Lyth,
Phys.\ Lett.\ B {\bf 526}, 173 (2002).

\bibitem{pyro}
R.~Kallosh, L.~Kofman and A.~D.~Linde,
Phys.\ Rev.\ D {\bf 64}, 123523 (2001).

\bibitem{lms}
H.~Liu, G.~Moore and N.~Seiberg,
gr-qc/0301001.

\bibitem{juan}
P.~Crotty, J.~Garc{\'\i}a-Bellido, J.~Lesgourgues and A.~Riazuelo,
astro-ph/0306286.

\bibitem{book}
A.~R.~Liddle and D.~H.~Lyth,
{\it Cosmological inflation and large-scale structure}
(Cambridge Univ. Press, Cambridge U.K., 2000).

\bibitem{laza}
G.~Lazarides,
Lect.\ Notes Phys.\  {\bf 592}, 351 (2002);
hep-ph/0204294;
hep-ph/9904502.

\bibitem{lw}
D.~H.~Lyth and D.~Wands,
Phys.\ Lett.\ B {\bf 524}, 5 (2002).

\bibitem{sylvia}
S.~Mollerach,
Phys.\ Rev.\ D {\bf 42}, 313 (1990).

\bibitem{lm}
A.~D.~Linde and V.~Mukhanov,
Phys.\ Rev.\ D {\bf 56}, 535 (1997).

\bibitem{dl}
K.~Dimopoulos and D.~H.~Lyth,
hep-ph/0209180.

\bibitem{luw}
D.~H.~Lyth, C.~Ungarelli and D.~Wands,
Phys.\ Rev.\ D {\bf 67}, 023503 (2003).

\bibitem{mt1}
T.~Moroi and T.~Takahashi,
Phys.\ Lett.\ B {\bf 522}, 215 (2001).

\bibitem{andrew}
N.~Bartolo and A.~R.~Liddle,
Phys.\ Rev.\ D {\bf 65}, 121301 (2002).

\bibitem{mt2}
T.~Moroi and T.~Takahashi,
Phys.\ Rev.\ D {\bf 66}, 063501 (2002).

\bibitem{fy}
M.~Fujii and T.~Yanagida,
Phys.\ Rev.\ D {\bf 66}, 123515 (2002).

\bibitem{hmy}
A.~Hebecker, J.~March-Russell and T.~Yanagida,
Phys.\ Lett.\ B {\bf 552}, 229 (2003).

\bibitem{hofmann}
R.~Hofmann,
hep-ph/0208267.

\bibitem{bck1}
M.~Bastero-Gil, V.~Di Clemente and S.~F.~King,
Phys.\ Rev.\ D {\bf 67}, 103516 (2003).

\bibitem{bck2}
M.~Bastero-Gil, V.~Di Clemente and S.~F.~King,
Phys.\ Rev.\ D {\bf 67}, 083504 (2003).

\bibitem{mormur}
T.~Moroi and H.~Murayama,
Phys.\ Lett.\ B {\bf 553}, 126 (2003).

\bibitem{ekm}
K.~Enqvist, S.~Kasuya and A.~Mazumdar,
Phys.\ Rev.\ Lett.\  {\bf 90}, 091302 (2003).

\bibitem{mwu}
K.~A.~Malik, D.~Wands and C.~Ungarelli,
Phys.\ Rev.\ D {\bf 67}, 063516 (2003).

\bibitem{postma}
M.~Postma,
Phys.\ Rev.\ D {\bf 67}, 063518 (2003).

\bibitem{fl}
B.~Feng and M.~Li,
Phys.\ Lett.\ B {\bf 564}, 169 (2003).

\bibitem{gl}
C.~Gordon and A.~Lewis,
Phys.\ Rev.\ D {\bf 67}, 123513 (2003).

\bibitem{kostas}
K.~Dimopoulos,
astro-ph/0212264.

\bibitem{giov}
M.~Giovannini,
Phys.\ Rev.\ D {\bf 67}, 123512 (2003).

\bibitem{lu}
A.~R.~Liddle and L.~A.~Ure\~{n}a-L\'{o}pez,
astro-ph/0302054.

\bibitem{mcdonald}
J.~McDonald,
hep-ph/0302222.

\bibitem{ejkm}
K.~Enqvist, A.~Jokinen, S.~Kasuya and A.~Mazumdar,
hep-ph/0303165.

\bibitem{dlnr}
K.~Dimopoulos, D.~H.~Lyth, A.~Notari and A.~Riotto,
J. High Energy Phys. {\bf 07}, 053 (2003).

\bibitem{ekm2}
M.~Endo, M.~Kawasaki and T.~Moroi,
hep-ph/0304126.

\bibitem{pm}
M.~Postma and A.~Mazumdar,
hep-ph/0304246.

\bibitem{kkt}
S.~Kasuya, M.~Kawasaki and F.~Takahashi,
hep-ph/0305134.

\bibitem{lw03}
D.~H.~Lyth and D.~Wands,
astro-ph/0306498.

\bibitem{lw032}
D.~H.~Lyth and D.~Wands,
astro-ph/0306500.

\bibitem{lm1}
D.~H.~Lyth and K.~Malik, in preparation.

\bibitem{note2}
Recently, a different  
alternative to the inflaton scenario  has been proposed 
in Refs.~\cite{dgz,kofman,emp},  in which some `modulating'
field perturbs the inflaton decay rate  without ever contributing
significantly to the energy density. The properties of the
modulating field have to be  much more special than those of the 
curvaton field.

\bibitem{dgz}
G.~Dvali, A.~Gruzinov and M.~Zaldarriaga,
astro-ph/
0303591.

\bibitem{kofman}
L.~Kofman,
astro-ph/0303614.

\bibitem{emp}
K.~Enqvist, A.~Mazumdar and M.~Postma,
Phys.\ Rev.\ D {\bf 67}, 121303 (2003).

\bibitem{randall}
T.~Moroi and L.~Randall,
Nucl.\ Phys.\ B {\bf 570}, 455 (2000).

\bibitem{randall2}
M.~Dine, L.~Randall and S.~Thomas,
Nucl.\ Phys.\ B {\bf 458}, 291 (1996);
Phys.\ Rev.\ Lett.\  {\bf 75}, 398 (1995).

\bibitem{thermal2}
D.~H.~Lyth and E.~D.~Stewart,
Phys.\ Rev.\ D {\bf 53}, 1784 (1996).

\bibitem{note3}
Note that, in view of Eq.~(\ref{rp}), $w_\sigma$ may be in principle
time-dependent.

\bibitem{turner}
M.~S.~Turner,
Phys.\ Rev.\ D {\bf 28}, 1243 (1983).

\bibitem{note4}
We ignore cubic $A$-terms.

\bibitem{note5}
Here, $g$ may also include loop factors 
or number of degrees of freedom etc.

\bibitem{note6}
For example, in the case of no-scale supergravity, an (approximate) Heisenberg 
symmetry suggests that \mbox{$c\sim 10^{-2}g^2$}.

\bibitem{note7}
Note that the full scalar potential includes also the contribution of the 
inflaton field, which, at least until reheating, is much larger than $V_0$.

\bibitem{note8}
Indeed, by retaining only the kinetic terms in Eq.~(\ref{KG}), one easily 
finds that \mbox{$\sigma=\sigma_0+(\dot{\sigma}_0/3H)[1-e^{-3H(t-t_0)}]$}, 
which means that the 
rapid roll due to the kinetic energy will cease in less than a Hubble time.

\bibitem{stoch}
A.~A.~Starobinsky,
Phys.\ Lett.\ B {\bf 117}, 175 (1982);
A.~D.~Linde,
{\em Particle Physics And Inflationary Cosmology},
(Harwood, Chur Switzerland, 1990).

\bibitem{note9}
This is easy to understand using the uncertainty principle (which 
governs quantum fluctuations) $\Delta E\cdot\Delta t = 1$, if one considers 
the energy within a Hubble volume
\mbox{$\Delta E\simeq\rho_{\rm qm}\times H_*^{-3}$} and 
the time necessary for horizon exit \mbox{$\Delta t\simeq H_*^{-1}$}.

\bibitem{note10}
This is the source of our disagreement with 
Ref.~\cite{postma}, where it is claimed that the border of the quantum regime 
is determined by comparing the energy density of the quantum fluctuations 
$\rho_{\rm qm}$ with the potential $V$ and not the kinetic energy density 
$\rho_{\rm kin}$ of the rolling scalar field. If one adopted such a view then 
there would be no quantum regime on top of local maxima (e.g. in the case of
eternal inflation) or at inflection points, where \mbox{$V'=0$}, as long as 
$V$ is large enough. Moreover, the fact that one can always add a cosmological 
constant would have really caused confusion since a change of $V$ would
have affected the location of the borders of the quantum regime.

\bibitem{note11}
Note that, since \mbox{$V(\sigma)/\rho_{\rm kin}=$ const.}, the
equation of state for the curvaton has a constant $w_\sigma$ both in the
oscillating case and in the case of the scaling solution.

\bibitem{note12}
This is 
true if $c$ does not change dramatically at the end of inflation. If, however, 
\mbox{$c\sim 1$}, at the end of inflation the field may find itself in the 
quasi-quadratic regime instead.

\bibitem{attr}
A.~R.~Liddle and R.~J.~Scherrer,
Phys.\ Rev.\ D {\bf 59}, 023509 (1999).

\bibitem{note13}
This is our main difference with Ref.~\cite{emp} where it is claimed 
that there is substantial damping of the field's perturbation between the end 
of inflation and the onset of the oscillations. The damping found in 
Ref.~\cite{emp} is due to the incorrect assumption that the field is 
``critically damped'', i.e. that \mbox{$V''\sim H^2$}, during this time 
interval, which would lead to substantial roll that reduces the 
perturbation.

\bibitem{note14}
Note that, if $c$ is positive during inflation, then 
\mbox{$\sigma_{\rm end}\ll\sigma_{\rm min}^{\rm end}$}. This, however, 
means that the oscillations will still have initial amplitude 
$\sim\sigma_{\rm min}^{\rm end}$ and so, for the amplitude of the oscillating 
field, Eq.~(\ref{fend-c}) is valid.

\bibitem{note15}
For example, in
the matter era the value of the field is described by the Eq.~(\ref{frz}) with 
the substitutions \mbox{$\delta\sigma\rightarrow\sigma$} and
\mbox{$\sqrt{V''_0}\rightarrow m$}.

\bibitem{KT}
E.~W.~Kolb and M.~S.~Turner,
{\it The Early Universe}
(Addison-Wesley Pub. Co., Redwood City U.S.A., 1990).

\bibitem{note16}
Note that the gravitino bound on \mbox{$T_{\rm reh}$} may be 
substantially relaxed by the extra entropy production if the curvaton decays 
after it dominates the Universe.

\end{thebibliography}
\end{document}